\documentclass[onecolumn, 12pt]{article}
\usepackage[english]{babel}

\usepackage[utf8]{inputenc}
\usepackage[pdftex]{graphicx}
\usepackage{authblk}
\usepackage{dcolumn}
\usepackage{amsmath}
\usepackage{longtable}
\usepackage{tabularx}
\usepackage{setspace}
\usepackage{multirow}
\usepackage{mathtools}
\usepackage{tikz}
\usepackage{lscape}
\usepackage{amssymb}
\usepackage{url}
\usepackage{arydshln}
\usepackage{graphicx}
\usepackage{lscape}
\usepackage{enumerate}
\usepackage{enumitem}

\usepackage{lineno}

\usepackage{titletoc}
\usepackage{titlesec}

\usepackage{pdfpages}

\usepackage[sorting=none]{biblatex}
\usepackage{csquotes}
\MakeOuterQuote{"}
\usepackage[hypertexnames=false]{hyperref}

\hypersetup{colorlinks=true, linkcolor=orange, citecolor=purple, urlcolor=blue}

\definecolor{applegreen}{rgb}{0.55, 0.71, 0}
\definecolor{wine}{RGB}{114,47,55}

\usepackage{lscape}

\usepackage[font=small]{caption}

\RequirePackage[twoside,letterpaper,includeheadfoot,layoutsize={8.125in,10.875in},layouthoffset=0.1875in,layoutvoffset=0.0625in,left=50pt,right=50pt,top=50pt,bottom=50pt,headheight=0pt,headsep=10pt,footskip=25pt]{geometry}
\makeatletter
\patchcmd{\maketitle}{\LARGE \@title}{\fontsize{16}{19.2}\selectfont\@title}{}{}
\makeatother

\begin{document}

% \linenumbers

\title{Structural Gender Imbalances\\[0.4cm] in Ballet Collaboration Networks}

\author[1]{\small Yessica Herrera-Guzm\'{a}n}
\author[2]{\small Eun Lee}
\author[3,4,*]{\small Heetae Kim}

\setstretch{0.8}

\affil[1]{Research Center for Social Complexity, Universidad del Desarrollo, Chile.} 

\affil[2]{Department of Scientific Computing, Pukyong National University, Republic of Korea.}

\affil[3]{Department of Energy Engineering, Korea Institute of Energy Technology, Republic of Korea.}

\affil[4]{Data Science Institute, Faculty of Engineering, Universidad del Desarrollo, Chile}

\affil[*]{Corresponding Author. Email: \href{hkim@kentech.ac.kr}{hkim@kentech.ac.kr}}

\date{}

\maketitle

\vspace{-30pt}
\begin{abstract}
\onehalfspacing

Ballet, a mainstream performing art predominantly associated with women, exhibits significant gender imbalances in leading positions. 
However, the collaboration's structural composition on gender representation in the field remains unexplored. 
Our study investigates the gendered labor force composition and collaboration patterns in ballet creations. 
Our findings reveal gender disparities in ballet creations aligned with gendered collaboration patterns and women occupying more peripheral network positions respect to men. 
Productivity disparities show women accessing 20-25\% of ballet creations compared to men. 
Mathematically derived perception errors show the underestimation of women artists' representation within ballet collaboration networks, potentially impacting women's careers in the field. 
Our study highlights the structural disadvantages that women face in ballet and emphasizes the need for a more inclusive and equal professional environment to improve the career development of women in the ballet industry. 
These insights contribute to a broader understanding of structural gender imbalances in artistic domains and can inform cultural organizations about potential affirmative actions towards a better representation of women leaders in ballet.

\end{abstract}

\clearpage

\setstretch{1}
\section{Introduction}

One broadly investigated complex socioeconomic problem, is global economic inequality~\cite{becker1985human, atkinson2015inequality, piketty2015capital}. 
There is growing evidence that economic inequality affects artists and, more importantly, women artists, to enjoy of economic growth and access to leading positions in their careers \cite{topaz2019diversity, lindemann2016asymmetrical}. 
It is therefore becoming increasingly important to understand the social dynamics of gender inequalities in the arts. 
%%%
In specific, ballet is widely recognized and appreciated around the world, and is assumed as a women-dominated profession~\cite{daly1987balanchine, homans2010apollo}. 
However, recent reports show considerable gender imbalances where men specifically dominate leading positions~\cite{defrank2016slow}. 
%%%%

The lack women's representation as leaders in ballet has been widely discussed within the dance community, claiming for more equal professional opportunities~\cite{gendergapballet, balletworld}.
For example, data from American dance companies reveals the unequal representation of women (less than 40\%) in artistic and executive positions~\cite{dance_data}, while the overall participation of women in the workforce is about 70\%~\cite{bureau_data}.
This difference of women's representation in leadership roles raises the question of whether or not women face a `glass ceiling' barrier in the ballet industry~\cite{cotter2001glass, williams1992glass}. 

In our complex society, individual characteristics ---such as race, religion, education, or gender--- have meaningful effects in social behaviors that shape structural disparities, which may be a result of homophily, the preference of individuals to connect with similar others~\cite{mcpherson2001birds}.
Network research reveals that structural properties influence the access to information~\cite{granovetter1973strength, pan2012strength}, creativity~\cite{uzzi_collaboration_2005}, productivity~\cite{abbasi2012egocentric}, and career success~\cite{juhasz2020brokering}. 
Moreover, homophilic behaviours embedded to an imbalanced social structure can negatively affect the ranking of individuals from minority groups by enhancing segregation effects~\cite{karimi2018homophily}.
In an imbalanced social structure, individuals may inaccurately estimate the frequency of the minority group, resulting in perception errors regarding the representation of attributes in a social network~\cite{festinger1954theory, lerman2016majority}. 
As a result, the importance of the minority group can be over or underestimated respect of what can be expected from the real representation in the network~\cite{lee2019homophily}. 
Since perception errors could reinforce unequal patterns in social connections, such as collaborations, understanding the role of network structure regarding gender imbalances can give an insight to an intervention of equal opportunities in professional positions. 

Despite the collaborative nature of ballet creations, previous reports have primarily focused on quantifying the percentage of women and men artists involved, while the role of collaboration structures in contributing to gender imbalances in ballet remains poorly understood.
The existing literature evidences that gendered variations in social network structure contribute to different professional outcomes for men and women~\cite{yang2019network}, highlighting the importance of investigating the gender representation in collaborative structures.
Yet, there is a lack of systematic studies exploring the representation of women and the structural properties of ballet's professional network.
%%%

In this work, we investigate the social network structure and collaboration patterns of ballet creations. 
We hypothesize that, if the network structure is unbalanced by gender, the imbalanced social structure will align with unequal collaborative behaviors and the existence of perception errors, which could explain why females do not undertake or are overlooked from leading positions in this industry. 
This research relies on the stable collaborative structure of ballet, which has remained largely unchanged since its origins, to conduct a network analysis with scientific validity.

We construct collaboration networks from four renowned ballet companies and analyze their gender composition. 
The collaboration structures studied here mainly comprise a core structure of ballet creators, such as choreographers, composers, and costumes and light designers.
We compare the real-world collaboration structures with randomized network models. 
We specifically explore the structural gendered differences and the labor force composition in highly central positions.
We also measure the formation of perception errors on the women's group to examine a possible relationship between gendered collaboration networks and perceived working environment.  
To the best of our knowledge, our study is the first attempt to understand the structural gender imbalances in major ballet companies.
This research will help understand the underlying social mechanisms driving gender inequalities in a highly collaborative performing art.
We hope that this work will shed light for more effective interventions to reduce the segregation of women in creative careers.

\section{Methods}

\subsection{Network of ballet creators}

We construct the collaboration networks of ballet creators from four major ballet companies ---the American Ballet Theatre (ABT)~\cite{ABT}, the New York City Ballet (NYCB) ~\cite{NYCB}, the National Ballet of Canada (NBC)~\cite{NBC}, and the Royal Ballet of the Royal Opera House (ROH)~\cite{ROH}--- based on their worldwide prestige and the availability of their historical repertoire in their website.
Company data are collected using a Robotic Process Automation method for web scraping \cite{van2018robotic}. 
Our data collection and research methods were approved on January 18th, 2023, by the Institutional Research Ethics Committee of Universidad del Desarrollo, in Chile. 

The collected data includes \textit{original ballet titles}, as stated in each company's repository, and refers to ballet works with \emph{artistic elements that remain constant across time, performances, and productions} (e.g. creators, libretto, music, genre).
When appropriate, ballet companies list revivals (recreated works), and/or company premieres (productions that were originally debuted at a different ballet company, but that are presented for the first time in the company listing the work).

Collaboration networks are formed from the teams of leading artists working together to create a ballet work. 
Teams of ballet creators are formed from each record of original ballet titles, which includes the credits of leading artists, such as principal creators (choreographer and composer) and specialized roles (librettist, costumes and lighting designer), and does not include the dancers or any other company members.
In a few occasions, companies report the producer, designer (unspecified), and media editor of a ballet work, and other team structures vary in size by adding multiple collaborators for the same role (e.g. two or more composers).
It is important to note that ballet is strongly recognized for its stable collaborative structure, comprising a core structure of leading artists, such as choreographer, composer, librettist, and costume and light designer.
Hence, in constructing the collaboration network of ballet creators, we consider all listed artists in each ballet title as equal contributors to the ballet creation.
Therefore, a \emph{ballet collaboration} is defined as the creative and collective effort between choreographers, composers, costume designers, lighting designers, and other artists listed by each company, for the creation of a ballet title.
For further details, please see Section~\ref{section:collaborations}. 

\begin{figure}[h!]
\centering
\caption{\textbf{Schematic representation of data processing and network construction.} 
    \textbf{a} The collected data contains ballet titles (the title of a ballet creation) and artists' names with their corresponding roles. 
    \textbf{b} The collected data is transformed to a bipartite network, where artists connect to ballet titles that they have been participated as creators. 
    \textbf{c} From the bipartite network, a projected unipartite network representing the collaboration of artists in a ballet company.} 
\includegraphics[scale=0.6]{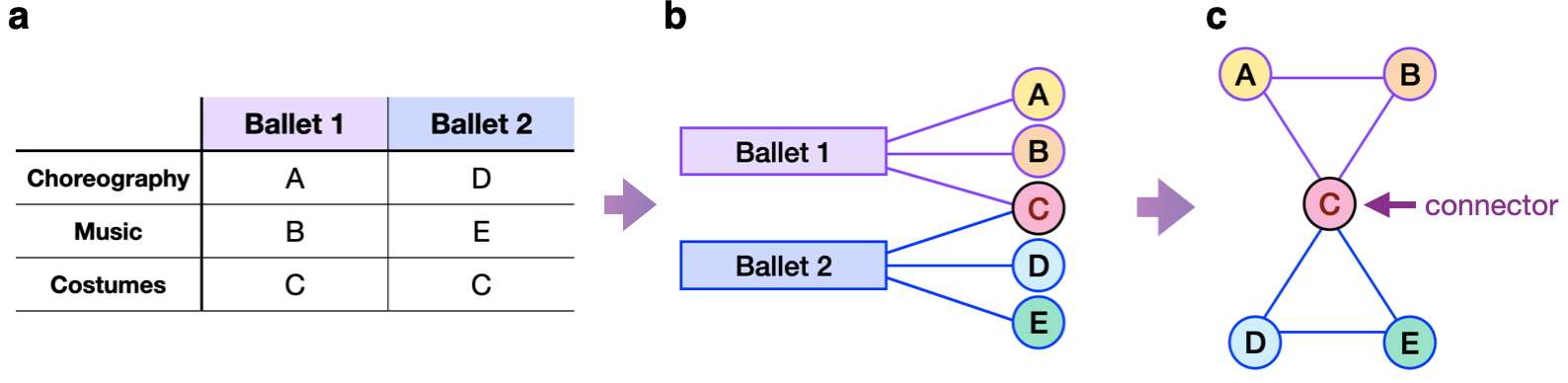}
\label{fig:model}
\end{figure}

The processing of the data is as follows. 
In Fig.~\ref{fig:model}a there is an illustration of the data showing a list of ballet titles (as an example, `Ballet 1' and `Ballet 2') with the names of ballet creators (A, B, C, D, and E), and their roles (e.g. Choreography, Music, and Costumes). 
Then, all artists who collaborated in a ballet creation together are part of the same team. 
To construct the collaboration network of each company, we first build a bipartite network between ballet creations and artists, as seen in Fig.~\ref{fig:model}b, where left nodes represent ballet titles and right nodes display the artists that created a ballet title.
Next, artists’ collaborations are projected to an undirected graph, as shown in Fig.~\ref{fig:model}c, where each node represents one artist, and a link between two artists denotes their collaboration in the same ballet creation.
An artist who teams up in more than one ballet creation will connect multiple artists in the same company, becoming a connector in the collaboration network. 
%%

%%%%%%%%%%%%%%%%%%%%%%%%%%%%%%%%%%%%%%%%%%%%%%%%%
%%%%%%%%%%%%%%%%%%%%%%%%%%%%%%%%%%%%%%%%%%%%%%%%%

The resulting empirical networks include about 300--560 ballet works, with a range of 490--850 artists (nodes) and 1900--3100 collaborations (links). 
In addition, he time of reported ballet creations ranges from 1930's to 2020's, making the networks comparable in terms of size and longevity. 
Basic network properties ---such as size of the giant component, average clustering coefficient~\cite{watts1998collective}, average shortest path~\cite{chen2008average}, and small-worldness~\cite{humphries2008network}--- can be seen in Table~\ref{tab:metrics}. 
%%

%%%%%%%%%%%%%%%%%%%%%%%%%%%
\begin{table}[h!]
\caption{\label{tab:metrics}Data description and basic network characteristics of the four ballet companies.}
\centering
\resizebox{\textwidth}{!}{\begin{tabular}{@{\extracolsep{5pt}} lcccc} 
\\[-1.8ex]\hline 
\hline \\[-1.8ex] 
 & \textbf{ABT} & \textbf{NYCB} & \textbf{NBC} & \textbf{ROH} \\ 
\hline \\[-1.8ex] 
Time window & 1940-2020 & 1935-2020 & 1951-2019 & 1931-2012 \\ 
No. of Ballets & 491 & 560 & 349 & 449 \\ 
Artists (nodes) & 779 & 549 & 579 & 847 \\
Collaborations (links) & 2323 & 2202 & 1913 & 3107 \\
Average degree & 5.9641 & 8.0219 & 6.6079 & 7.3365 \\ 
Density & 0.0076 & 0.0146 & 0.0114 & 0.0086 \\ 
Clustering coefficient  $C_{avg}$ & 0.7556 & 0.7710 & 0.7776 & 0.8033 \\
Average Path Length $L$ & 3.52 & 2.63 & 3.60 & 3.49 \\
Small-worldness, $S$ & 97.73 & 67.57 & 75.07 & 102.99 \\
\textbf{Giant Component} &  &  & \\
\hspace{2mm} Artists (as percentage) & 715 (91.78) & 643 (98.90) & 540 (93.26) & 816 (96.34) \\ 
\hspace{2mm} Collaborations (as percentage) & 2251 (96.90) & 2195 (99.68) & 1863 (97.38) & 3071 (98.84) \\ 
\hspace{2mm} Density & 0.0088 & 0.0149 & 0.0128 & 0.0092 \\
\hspace{2mm} Diameter & 8 & 5 & 10 & 8 \\
\hline \\[-1.8ex] 
\end{tabular}}
\end{table}
%%%%%%%%%%%%%%%%%%%%%%%

\subsubsection{Gender inference}

Artists' names were processed for misspelling, middle names, and initials to distinguish artists' identity.
The names are held constant if reported across multiple companies. 
Then, we infer artists' gender by using \texttt{gender} package for R~\cite{karimi2016inferring, blevins2015jane}.
This package contains names from various countries and periods, and infer names from standardized databases (\texttt{ssa}, \texttt{ipums}, \texttt{napp}, and \texttt{demo}), making it adequate for this study since the collected data contains names of artists with diverse nationalities and were born in the 19th and 20th centuries.

To estimate an artist's birth year, we assume that each artist was at least 20 years old when they participated in a ballet creation for the first time.
Thus we subtract 20 years from the year of the first ballet production of an artist in our data as a proxy of the minimum age for a productive life in ballet. 
This method considers a range of 10 years ($\pm 5$ years from the estimated birth date).
Then, the \texttt{gender} package estimates a probability that a person would have certain gender with the name.
If the probability is larger than or equal to $0.7$, the corresponding gender is assigned to each artist. 
Here, the assigned `gender' is a binary property (Woman, Man) and does not consider other gender assignments. 
Note that the inferred gender does not refer directly the sex of the artist nor the self-assigned gender chosen by each artist., but is used as estimate of the social construction of gender. 
The names which were not able to assign gender with this method were manually assigned after a web search of the artist's identity.

\subsection{Network analysis and gendered differences in centrality}

We measure four network metrics to understand the importance or centrality of artists in the collaboration networks: 
\begin{enumerate}
    \item Degree centrality is computed following ~\cite{freeman1978centrality} to measure the number of total connections of a node.
    This metric can capture the individual access to a richer social capital.
    \item Harmonic centrality is computed following~\cite{boldi2014axioms}, and is a variant of closeness centrality created to deal with unconnected graphs to measure the distance one node has respect to all other nodes in the network.
    In other words, harmonic centrality captures the position of nodes to efficiently reach distant parts of the network.
    \item Betweenness centrality is computed considering all pair of nodes as described in~\cite{freeman1977set} to measure the number of shortest paths between two pairs of nodes that pass through a node in a network.
    This metric captures what nodes are best intermediaries or bridges between different parts of the network. 
    \item Eigenvector centrality is computed following~\cite{bonacich2007some} and measures the importance of a node based on the centrality of its neighboring nodes.
    This centrality informs about the nodes who are connected to other influential or central nodes, as these can help gain social prestige in the network. 
\end{enumerate}

These metrics are informative on the differential ranking of individuals embedded in the network~\cite{wasserman1994}.
For example, one artist with high centrality (e.g. degree) should indicate that the artist has multiple connections in the network, then being well positioned to have more access to information, social connections, and professional opportunities. 
In a global sense, these centrality metrics help identify structural patterns within a network, providing insights into the underlying relationships between individuals that ultimately shape the network. 
%%%

%%%
From the centrality metrics, we sort all artists by their centrality in descending order, and selected a group of top 20 artists, referred as \emph{Top-Central Artists} (TCA) in this study. 
Let consider the ranking of centralities $C(r)$, where $C$ denotes a corresponding centrality value of an artist at a given rank $r$, for $r = \{1, 2, \dots, 19, 20 \}$, so $r = 1$ represents the most central artist having the highest corresponding centrality (e.g. $C(1) =  0.8$), and $r=20$ will have the lowest centrality (e.g. $C(20) =  0.2$). 
We select the top 20 as this fraction captures the largest observed variation of centrality values in the empirical networks, and between and within gender groups. 
Then, by analyzing the TCA, we capture the artists with best connected individuals in the network and the differences in network positions across gender categories.
%%%%

%
Next, we implement the TCA ranking to form three independent groups: the first group is for all artists in a company's collaboration network, labeled as $\text{TCA}_{\text{Network}}$, and the other two groups are for a company's artists grouped by gender, which results in two separate rankings for $\text{TCA}_{\text{Women}}$ and $\text{TCA}_{\text{Men}}$.
All centralities are then normalized by the maximum value of the centrality within company group (Network) and by company gender groups (Women, Men), to have a linear scaling of $[0,1]$ range. 
In more detail, the $\text{TCA}_{\text{Network}}$ uses a dense rank function, which generates rank ties for observations with the same centrality values, so a variation of the total number of artists is possible if there are artists with equal centrality at each rank.
For $\text{TCA}_{\text{Women}}$ and $\text{TCA}_{\text{Men}}$, the tied centrality is not considered to keep an equivalent number of women and men artists (i.e. 20 artists per group).

Separately, we quantify the women ratio $R_{\text{Women}}$ in each $\text{TCA}_{Network}$, computed as $R_{{\text{Women}}} = \frac{\sum_i^{N_{\mathrm{TCA}}} \theta(i)}{N_{\mathrm{TCA}}}$.
Here, $i$ denotes an index for an artist who is in a corresponding TCA, where $N_{\mathrm{TCA}}$ represents the total number of artists in a $\text{TCA}_{Network}$, and $\theta(i)=1$ when an artist is woman, or $0$ for men. 
Then, $R_{\text{Women}}$ provides the fraction of women artists who belong to the group of best connected individuals in the collaboration network of a ballet company. 
A numerical fraction of women artists at the network level of $0.5$ is assumed as a gender-balanced collaborations, and we call this situation as `neutral' composition.
%%%

Further, we measure the difference in centrality $\Delta C(r)$ between two rank-matched artists from each gender group is measured as $\Delta C(r) = C_{\text{Men},r} - C_{\text{Women}, r}$. 
Here, each woman artist from $\text{TCA}_{\text{Women}}$ is matched to their corresponding $r$ pair from $\text{TCA}_{\text{Men}}$.  
That is to say, if there is a woman artist ranked $1$ in $\text{TCA}_{\text{Women}}$ with a centrality value of $0.4$, she is at the most central position in the women's group, and it can be written as $C_{\text{Women, 1}} = 0.4$. 
The counterpart of man artist, who is ranked $1$ as well in $\text{TCA}_{\text{Men}}$ will be $C_{\text{Men, 1}} = 0.5$, if he has a centrality of $0.5$. 
Then, $\Delta C(1) = 0.5-0.4 = 0.1$.
If $\Delta C(r) > 0$, it means that a man artist is located on more central position than the woman counterpart. 
%%%%

\subsubsection{Null model analysis}

We compute two different null models by simulating 100 synthetic networks derived from the representation of each company's empirical collaboration network. 
With the help of the null models, we remove the collaborator- or gender-preferences by shuffling collaborations (links) or artists' attributes (gender) in the collaboration network.
The overall purpose of the null models is create a baseline of randomly created networks, which would allow us to determine the absence or existence of randomness in the observed patterns respect to the empirical network.

\begin{enumerate}[label=(\arabic*)]
    \item Edge-shuffled model: In this model, edges are randomly rearranged in the network while preserving artists' degrees.
    This means that the total number of collaborations per artists are preserved, as well as the total number of artists (nodes) in the network and artists' gender. 
    We use the ‘random\_reference’ function of \texttt{NetworkX}~\cite{maslov2002specificity}. 
    From this shuffling, we remove the gendered correlation from empirical collaboration networks.
    Therefore, the resulting synthetic networks show collaboration structures when there is no gender preference.

    \item Gender-shuffled model: This model shuffles the gender of artists while holding all network properties constant. 
    Here, the empirical network structure is used as a reference, without nodes’ attributes, over which a dictionary containing the gender of all nodes is used to randomly assign artists' gender, while preserving the real fraction of women and men in the network.
    In this way, artists’ network position are preserved, but their gender is randomized in each iteration.
    Therefore, the resultant networks display an artificial collaboration pattern without a correlation between an artist's gender and position, as well as a gendered collaboration assortativity.
\end{enumerate}

To test a null hypothesis distribution, we compute the Z-score for a distinction between the centrality values from the empirical networks and those from the null models.
We denote the observed centrality by rank in the real network as $C(r)_{\text{real}}$, and that of the null model as $C(r)_{\text{null}}$.
Then, the $Z$-score for any TCA group uses the centrality from the empirical network, $C(r)_{\text{real}}$ , and the averaged centrality of 100 null models, $\bar{C}(r)_{\text{null}}$, so that $Z(C) = \frac{C(r)_{\text{real}} - \bar{C}(r)_{\text{null}}}{\sigma(C(r)_{\text{null}})}$.
Z-score of $\Delta C(r)$ in the empirical network is also measured with the values of the synthetic networks as $Z(\Delta C) = \frac{\Delta C(r)_{\text{real}} - {\Delta \bar {C}}(r)_{\text{null}}}{\sigma(\Delta C(r)_{\text{null}}})$. 
%%%

\subsection{Perception error on women artists}

To understand more about the implications of the gendered differences in the collaborative environment, we use a mathematical approach to measure the existence of perception errors based on~\cite{lee2019homophily}.
Perception errors refer to the inaccuracy in the estimation of the frequency of an attribute ---usually of a minority group--- in a social network, perceived from the frequency of that attribute within the individual local network~\cite{lerman2016majority, karimi2018homophily}. 
In this research, perception errors are the difference in the perceived fraction of women artists from the local network, respect to the fraction in the entire network. 
For instance, if in the local network there are mostly women, but in the entire network there are more men, one individual will have a perception error above one that over estimates the size of the women’s group, while the opposite happens for the under-estimation of the women's group, with a value below 1.
Thus when the perception error is equal to 1, that means that the perception of the fraction of women in the network is accurate. 
The perception error $B$ of an individual artist $i$ is thus computed as $B_{i} = \frac{W_{i}}{R_{\text{Women}}}$, where $W_{i}$ denotes the local fraction of women among $i$'s collaborators, and $R_{\text{Women}}$ refers to the real fraction of women in the network, as noted above.

Based on the individual artist's perception error, we measure an averaged perception error by gender group at a network-level, so $\bar{B}_{Women, Men} =\frac{\sum_i B_i}{N_{Network}}$, where $N_{Network}$ represents the total number of artists in a ballet company. 
Consequently, when $\bar{B} = 1$, it means that the overall perception of women on a company is accurate on average, and when $\bar{B} < 1 (\bar{B} >1 )$, a gender group underestimates (overestimates) the ratio of women artists on average. 
In addition, a gendered homophily is measured following the method in~\cite{lee2019homophily} to see gendered preferences of the collaboration networks. 
%%%

\section{Results}

Based on previous reports on the lack of representation of women in leading positions in ballet~\cite{dance_data}, we explore the general composition of the collaboration networks of ballet creators and the existence of gendered collaboration patterns in the professional environment. 
We also look into the composition of the most central network positions and the gender gap between men and women's centrality in the network; in addition, we measure the existence of perception errors of the women's artists group within ballet companies.
We compare network position and perception errors from the empirical network structures with two null model analyses.
%%%%

\subsection{Team structure and collaboration patterns}

\begin{figure}[!ht]
 \hspace*{-4mm}
    \caption{\textbf{Team composition and collaboration patterns by gender.} 
    Sample of the Royal Opera House (ROH). 
    Panel \textbf{a} shows the probability of a gender ratio in teams: more than 50\% of teams are composed only by men, while teams having only women are nonexistent. 
    Panel \textbf{b} shows the normalized frequency of same-gender collaborations in a team: women mostly collaborate alone in a men-dominated teams, while men collaborate more with other 3-5 men and form larger teams. 
    And panel \textbf{c} shows the number of ballet creations for each artist. 
    Productivity varies by gender, with less productivity for women artists. 
    Fit line by gender with 95\% confidence intervals.}
\includegraphics[width = 0.95\textwidth]{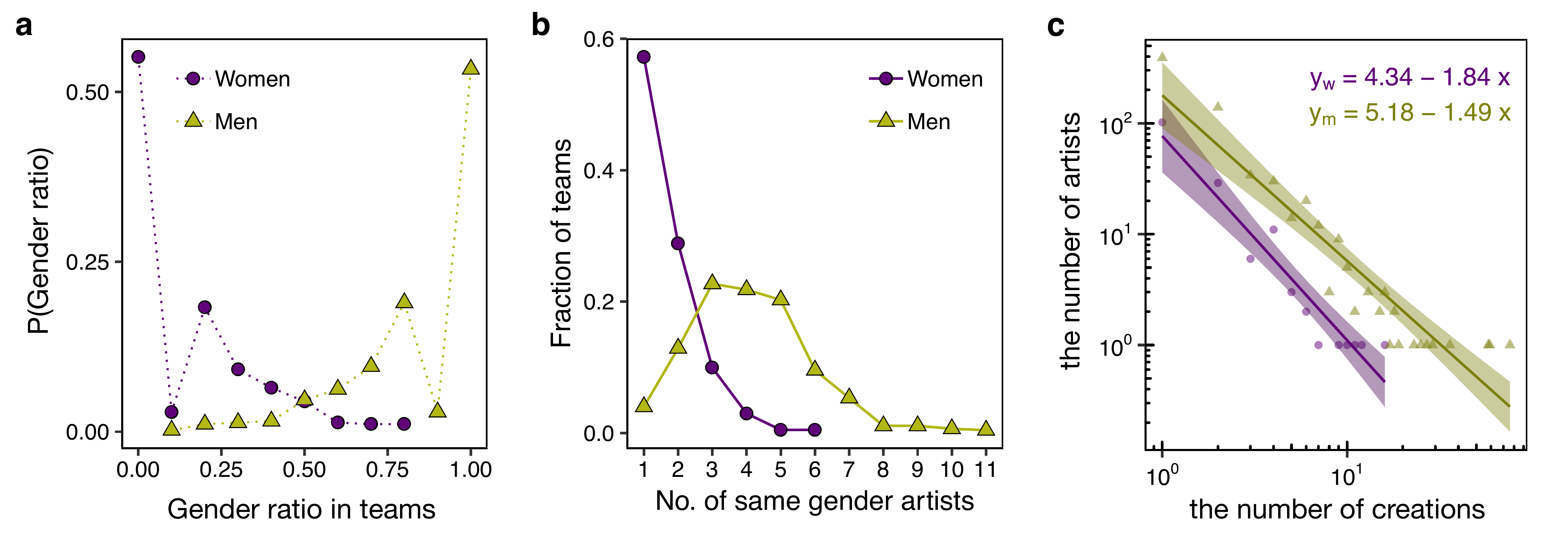}
\label{fig:collab_behav}
\end{figure}

% Teams
The most common team size for a ballet creation across companies is three to four (20--40\%), followed by five members (20\%), as shown in Fig.~\ref{Sfig_supp:teamsize}a.
This evidences that teams of ballet creators are mostly formed by the typical collaborative structure of leading artists. 
Fig.~\ref{fig:collab_behav}a shows a sample of the representation of women in a ballet company (ROH), shows that there are about 50\% of teams having 100\% men artists, and less than 10\% of teams have gender neutral ratio of 50\%.
Conversely, the majority of teams are composed with less than 50\% of women artists, regardless of their sizes, and teams having 100\% women artists is almost zero. 
%

% Roles
Dance communities have specifically reported an overlooking of women in choreographic leads, and our results suggest that women are less represented than men in general leading roles.
Exploring the team composition by artistic role, Fig.~\ref{Sfig_supp:women_artist} shows that the proportion of women is considerably low for Choreographer, Librettist, and Composer.
Other positions such as Costumes, Lighting, and Design have a relatively larger participation of women.
These results suggest variations between women and men regarding artistic roles are possible.
However, because most teams of ballet creators are formed by a core structure of leading artists, here we focus on the structural representation of women at both team and network-levels, rather than an individual-artistic role level.
%%%

% Collaborations
Further, in Fig.~\ref{fig:collab_behav}b we see that when women collaborate in one team, the frequency of working with other women in the same team is actually very low ($< 30\%$). 
These results describe that women artists mostly work in men-dominated environments.
In addition, men-alone teams are rather rare ($< 10\%$), as they tend to collaborate with at least other three to five men ($> 20\%$) and participate in considerably larger teams than women (up to 11 men in one team, at ROH). 
%

%%%%%%%%%%%%%%%%%%%%%%%%%%%%%%%%%%%%%%%%%%%%%%%%%
%%%%%%%%%%%%%%%%%%%%%%%%%%%%%%%%%%%%%%%%%%%%%%%%%%%%%

% Productivity
In terms of productivity, women artists are less involved in ballet creations than men artists.
One ballet creation refers to the participation of an artist in a team as leading artist for the creation of a ballet work. 
In NYCB and ROH, the most productive woman participates in about 20-25\% of the creations of the most productive man artist collaborated (ROH's maximum collaborations: Men = 76, Women = 16; NYCB's maximum collaborations: Men = 211, Women = 54, see Figures~\ref{fig:collab_behav}c, and ~\ref{Sfig_supp:collaborations}b). 
For NBC, the highest productivity is a bit similar for both genders. 
Women artists' highest productivity is just 86\% of the most productive man (NBC's maximum collaborations: Men=38, Women=33). 
Only at the ABT, the most productive woman artist exceeded in 20 collaborations to the most productive man artist (ABT's maximum collaborations: Men=35, Women=55).
Despite the exception, most women artists are less productive than their men counterparts, and the global picture for women is to work in men-dominated creative environments.
Team structures, and collaboration and productivity patterns, are similar across all companies studied here (for more details and figures by company, see Section~\ref{section:collaborations}).  
%%%

\subsection{Centrality differences by gender}

\begin{figure}[ht!]
\centering
\caption{\textbf{Distribution of artists and their collaborations.} Sample of ROH's collaboration network. Panel \textbf{a} shows nodes colored in purple/yellow for women/men. 
Node size proportional to degree centrality. 
The dyadic collaborations by gender are shown in: panel \textbf{b} for man-man collaborations; and panel \textbf{c} for woman-woman collaborations.
We see that women are visually less central than men, and their collaboration with other women are scarce and peripheral.}
\includegraphics[scale=0.39]{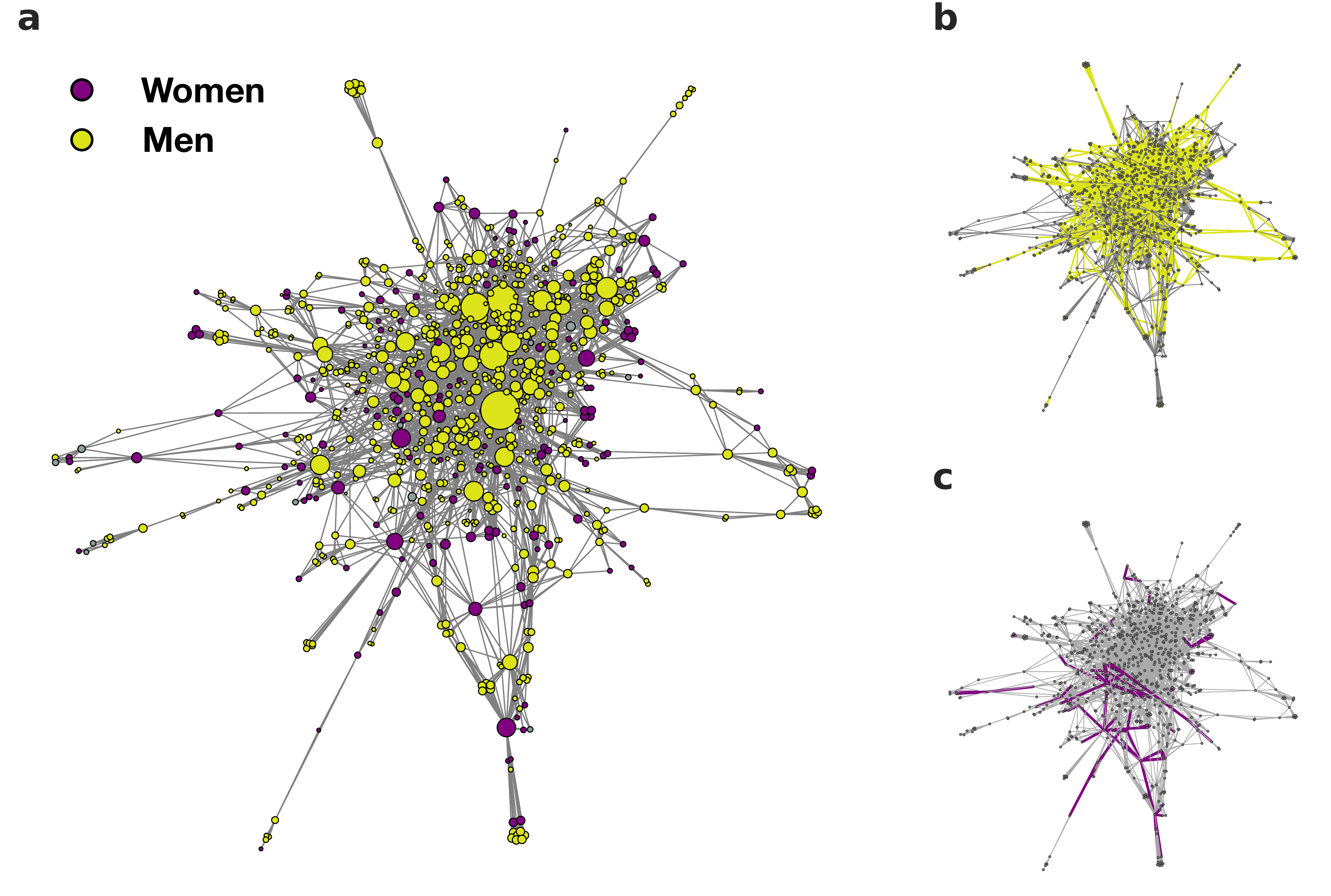}
\label{fig:network_sample}
\end{figure}

So far, we have observed a less frequent participation of women respect to men in ballet collaborations.
These observations raise the question: Does the low participation of women relate to their network position in the company?
To answer this question, we explore the distribution of artists' collaborations in the network. 
We first compute the fraction of women in the network, $R_{\text{Women}}$, and the proportion of dyadic interactions (see Table~\ref{tab:summary}), showing that most companies only have about $20\%$ of women in leading positions.  
Figure~\ref{fig:network_sample}a shows a network sample, where men (in yellow) are not only a majority but also with higher connectivity respect to women (in purple).
(See all companies' collaboration networks in~\ref{Sfig_supp:networks}). 
Moreover, the man-man connections are more than 60\% across companies (yellow links, Fig.~\ref{fig:network_sample}b) and mixed connections are about 30\% on average. 
On the other hand, woman-woman connections are less than 5\% of the total dyadic interactions (purple links, Fig.~\ref{fig:network_sample}c).
These results inform that, for every 4 men, there is only one woman in the network, a collaborative structure in which men artists are densely co-worked with other artists regardless of gender, locating at the center of the collaboration network, while women artists are sparsely distributed in the periphery of the network.  
%%%%%

%%%%%%%%%%%%%%%%%%%%%%%%%%%%%%%%%%%%%
%%%%%%%%%%%%%%%%%%%%%%%%%%%%%%%%%%%%%

%%%%%%%%%%%%%%%%%%%%%%%%%%%%%%%%%%%%%
\begin{table}[ht!]
\centering
\caption{\label{tab:summary}Network composition by gender. 
$R_{\text{Women}}$ is the fraction of women in a collaboration network. 
For the collaborations, the number of woman-woman/man-man/mixed dyadic interactions is counted.}
\begin{tabular}{@{\extracolsep{5pt}} lcccc} 
\\[-1.8ex]\hline 
\hline \\[-1.8ex] 
 & \textbf{ABT} & \textbf{NYCB} & \textbf{NBC} & \textbf{ROH} \\ 
\hline \\[-1.8ex] 
\textbf{Artists (nodes)} & 779 & 549 & 579 & 847 \\
\hspace{2mm} $R_{\text{Women}}$ & 0.22 & 0.19 & 0.18 & 0.19 \\
\textbf{Collaborations (links)} & 2589 & 2317 & 1956 & 3385 \\
\hspace{2mm} Woman-woman & 138 (5\%) & 64 (3\%) & 55 (3\%) & 135 (4\%) \\
\hspace{2mm} Man-man & 1603 (62\%) & 1576 (68\%) & 1336 (68\%) & 2353 (70\%) \\
\hspace{2mm} Mixed & 848 (33\%) & 667 (29\%) & 656 (29\%) & 897 (26\%) \\
\hline \\[-1.8ex] 
\end{tabular} 
\end{table}
%%%%%%%%%%%%%%%%%%%%%%%%%%%%%%%%%%%%%

\begin{figure}[h!]
\centering
\caption{\textbf{Fraction of women in $\text{TCA}_{\text{Network}}$ from the empirical network and null models.} 
Sample of ROH.
Average $R_{\text{Women}}$ of $\text{TCA}_{\text{Network}}$ in Edge-shuffled and Gender-shuffled models are shown with standard deviation. 
The null models show a fairer representation of women artists than the empirical network.}
\includegraphics[width = 0.8\textwidth]{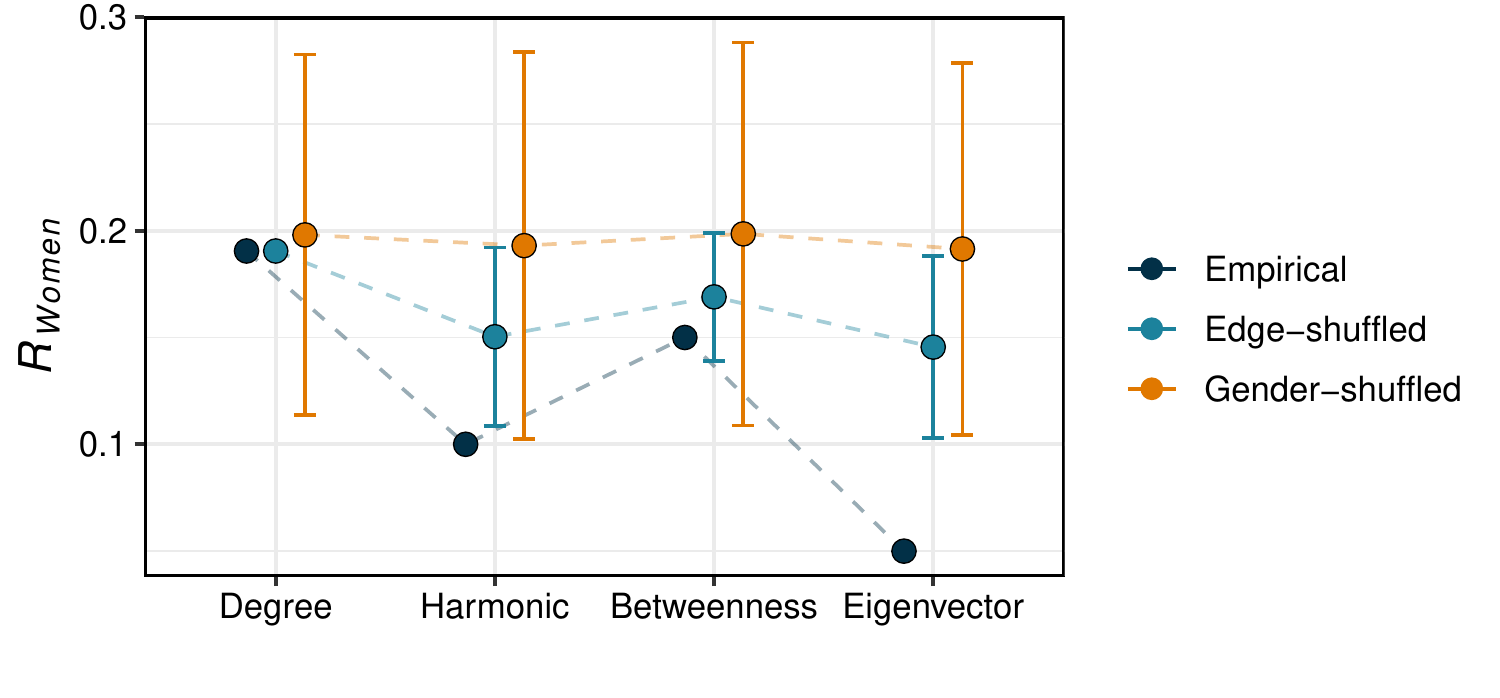}
\label{fig:women_fractionROH}
\end{figure}

We then evaluate the proportion of women in the group of top-central artists, $\text{TCA}_{\text{Network}}$, by their network centrality rank, $C(r)$, and observe that most companies have a lower representation of women respect to $R_{\text{Women}}$ in the empirical network. 
We observe an overall increase of $R_{\text{Women}}$ in the randomized models for all centralities (see all companies in Fig.~\ref{Sfig_supp:women_fraction}).
For example, the Edge-shuffled model improves $R_{\text{Women}}$ for harmonic centrality from $10\%$ to $15\%$, and Gender-shuffled model raises it up to $19\%$ in the sample of the ROH, a fraction that matches the $R_{\text{Women}}$ of the total empirical network (Fig.~\ref{fig:women_fractionROH}). 
Note that the Edge-shuffled model keeps $R_{\text{Women}}$ in TCA regarding degree centrality because the number of collaborations (degree) for an artist and their inferred gender are held constant in this model. 
These results suggest that the low representation of women artists in ballet creations could be related to gender assortative collaborations, and the current level of women artists' centrality is not a deterministic outcome of the small fraction of women artists in the company.  
That is to say, even when the fraction of women remains small in a network, women artists' representation could be improved if more equal collaborations for women were encouraged.
%%%

\begin{figure*}[ht!]
\centering
\caption{\textbf{Comparison of centralities for artists in $\text{TCA}_{\text{Women}}$ and $\text{TCA}_{\text{Men}}$ with null models}. 
These figures are for ROH's collaboration network.
Z-score of centralities ($Z(C)$) compared with \textbf{a} Edge-shuffled model and \textbf{b} Gender-shuffled model. 
The red line corresponds to the theoretical mean obtained from the null models indicating no difference. 
Panel \textbf{c} shows the gender gap in the degree centrality (normalized) between $\text{TCA}_{\text{Women}}$ and $\text{TCA}_{\text{Men}}$, revealing that $\text{TCA}_{\text{Men}}$ have higher degree centrality than their women counterparts. 
Panel \textbf{d} shows the distribution of $Z(\Delta C)$ separated by null model.}
\begin{tabular}{cc}
\includegraphics[width = 0.44\textwidth]{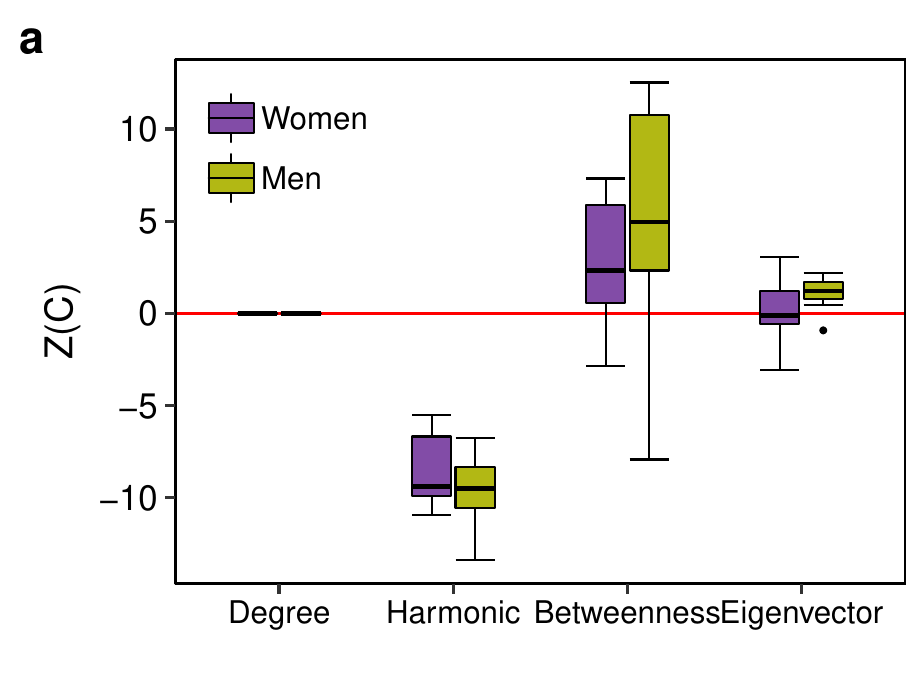}
\includegraphics[width = 0.44\textwidth]{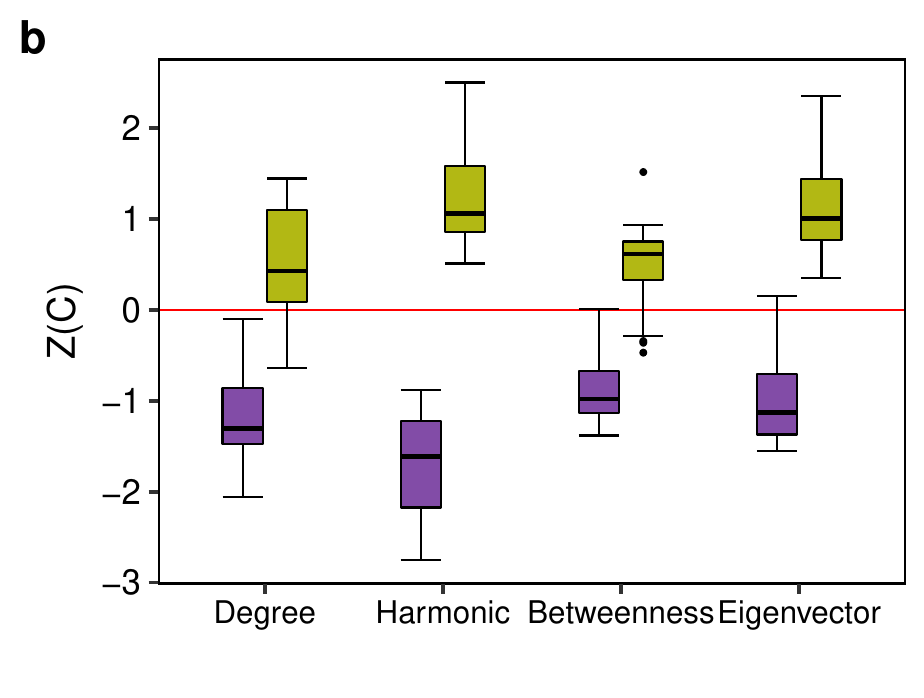}\\
\includegraphics[width = 0.36\textwidth]{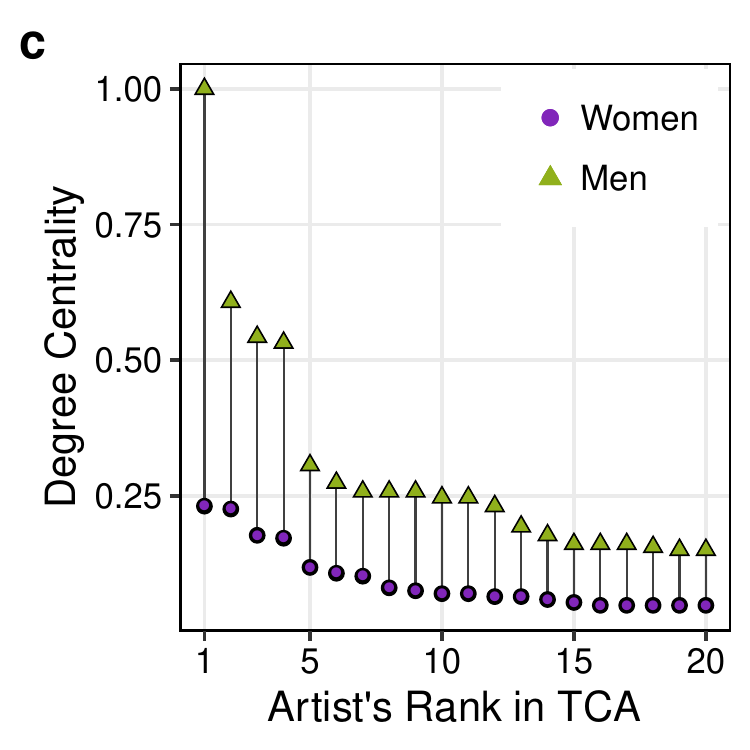}
\hspace{0.4cm}
\includegraphics[width = 0.47 \textwidth]{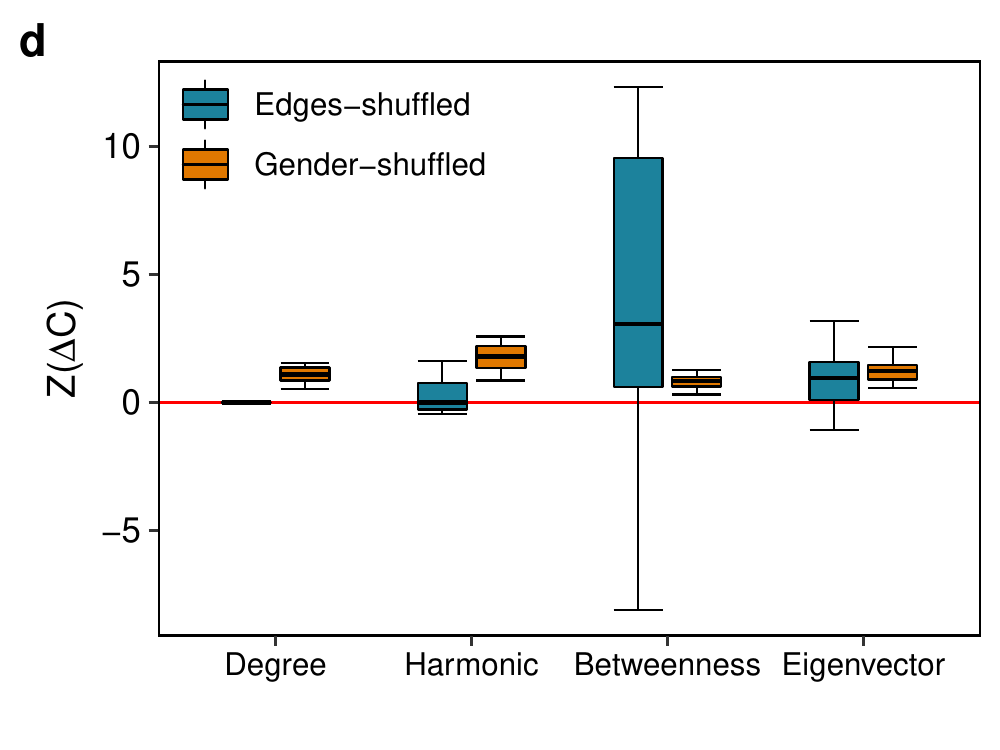}
\end{tabular}
\label{fig:zscore_ROH}
\end{figure*}

The $Z(C)$ reveals a general change in artists' centrality with the null models (see ROH's sample in Fig.~\ref{fig:zscore_ROH}, all companies in Fig.~\ref{Sfig_supp:zscore}a). 
For the Edge-shuffled model, only the harmonic centrality displays a negative Z-score for both women and men (sample in Fig.~\ref{Sfig_supp:zscore}a).
Harmonic centrality denotes an extent of an artist's closeness to other artists on average, so small value represents far distance between artists. 
The negative $Z$-score suggests that the distance among artists in empirical collaboration networks falls apart farther than the expected distance from the null models.
In other words, the distribution of TCA in the empirical networks is more central, suggesting that TCA can more efficiently reach other artists in the network respect to the distance observed in the null models. 
For the Gender-shuffled model, the negative women artists' $Z$-scores indicate that their positional importance can be improved in a synthetic network with collaboration imbalances (sample in Fig.~\ref{Sfig_supp:zscore}b). 
Altogether, our results suggest that differences in centrality among TCA may not be derived by random factors, but there may be underlying systematic social behaviors limiting women artists' collaborations and network position, regardless of their small fraction in the network.
%%%%

%%%%%%%%%%%%%%%%%%%%%%%%%%%%%%%%%%%%%
%%%%%%%%%%%%%%%%%%%%%%%%%%%%%%%%%%%%%

The difference in degree centrality ($\Delta C$) highlights that a man artist locates at more central position than the same-ranked women artist in her TCA group. 
Figure~\ref{fig:zscore_ROH}c shows a sample for degree centrality and reveals that the most central man is considerably more central than the most central woman. 
This trend of $\Delta C$ is observed across centralities with slight variations, confirming that men are considerably better positioned respect to women across companies (see gender gap in centrality for all companies in Fig.~\ref{Sfig_supp:centrality_gap}). 
Interestingly, all empirical $Z$-scores for $\Delta C$ are several standard deviations away compared to the null models (see all companies in Fig.~\ref{Sfig_supp:zscore}b). 
Figure~\ref{fig:zscore_ROH}c illustrates the variations by null model, and showcases that a large gender gap is less likely observed when the gender preference (Edge-shuffled) and gendered productivity correlation (Gender-shuffled) are destroyed. 
%%%%%

%%%%%%%%%%%%%%%%%%%%%%%%%%%%%%%%%%%%%%%%%%%%%%%
%%%%%%%%%%%%%%%%%%%%%%%%%%%%%%%%%%%%%%%%%%%%%%%%

\subsection{Perception error of women artists}

Given the observed structural imbalances in the ballet collaboration networks, the low participation of women in professional collaborations could affect the perceived frequency on women artists in the entire network. 
Perception errors are the distorted frequency estimation of an attribute in a social network by the individual local environment~\cite{lerman2016majority, karimi2018homophily}. 
Here, the perception error is defined as the fraction of the observed frequency of women in an artist's local collaboration network over the real fraction of women in the global network (see Methods).
That is, the perception error denotes a relative difference of women artists in the local collaboration environment of each artist and the actual women artists' frequency in each ballet company. 
From the individual-level perception error $B_i$, a gender group-level error $\bar{B}$ compares the average perception error for women and men. 
If $\bar{B} > 1.0$ ($\bar{B} < 1.0$), it means a gender group overestimates (underestimates) the global frequency of women artist. 
When $\bar{B} = 1.0$, it denotes an accurate perception on the women frequency (see Methods). 
We complement perception error with a measure of homophily. 
%%%%

%%%%%%%%%%%%%%%%%%%%%%%%%%%%%%%%%%%%%%%%%%%%%%%%
\begin{figure}[ht!]
\caption{\textbf{Average perception error $\bar{B}$ for ballet companies.} 
Average perception error by gender in each company, compared with those from null models. 
\textbf{a} ABT, \textbf{b} NYCB, \textbf{c} NBC, \textbf{d} ROH. 
Red line indicates $\bar{B} = 1$, an accurate perception of the women's group size. 
Line segments in gray guides the difference in perception error by gender group. 
Most gender groups misconceive the real fraction of women in their network, while the difference is reduced in Edge-shuffled model. 
The perceived frequency of women is considerably more accurate in the Gender-shuffled model.}
\centering
\includegraphics[width=0.97\textwidth]{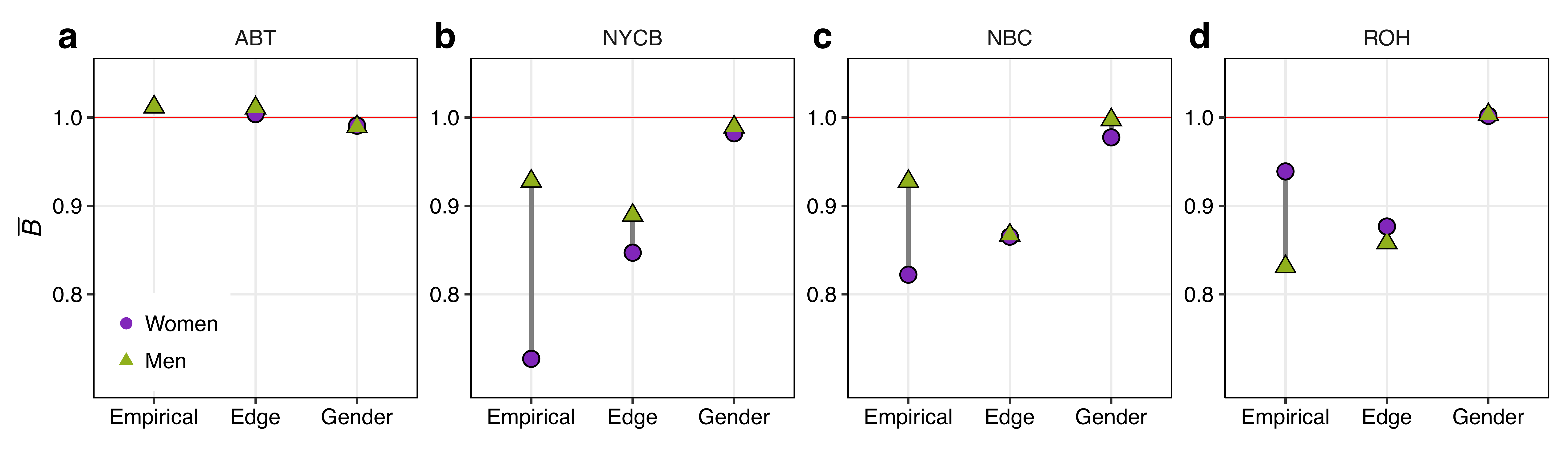}
\label{fig:perception}
\end{figure}
%%%%%%%%%%%%%%%%%%%%%%%%%%%%%%%%%%%%%%%%%%%%%%%%

Our results show that the empirical collaboration networks of the ABT, NYCB, NBC, and ROH, the women (men) artists' homophily is 0.56 (0.53), 0.47 (0.55), 0.45 (0.57), and 0.56 (0.63), respectively (1 is a perfect homophily, and 0 is a perfect heterophily situation). 
The ABT has relatively gender-mixed environment, resulting in both gender groups having relatively accurate perception on the global fraction of women artists, as shown in Fig.~\ref{fig:perception}a. 
Conversely, the rest of the companies demonstrate a wide difference in the perception error by gender, shown in Fig.~\ref{fig:perception}b--d. 
For instance, NYCB's men group underestimates women artists about 7\%, but their women group underestimates themselves about 27\%, showing a 20\% difference in the perception of women between the two groups.
Such difference may be related with men artists' strong homophily in NYCB collaborations, and women artists' gender-heterophilic collaborations (woman-man heterophily $0.53 >$, and woman-woman homophily $0.47$), and indicates a perceived underestimation of women artists by themselves. 
In ROH, women artists have a more accurate estimation of women artists respect to men artists, which aligns with their collaborative behaviors, where women artists collaborate more with other women artists than other men artists (woman-man heterophily $0.44 <$ woman-woman homophily $0.56$). 
Yet, the difference in perception still exists since men artists collaborate mostly with men artists (man-man homophily $0.63$, man-woman heterophily $0.37$), and the assortative collaboration widens the difference in perception between gender groups. 
%%%%%

Interestingly, the Edge-shuffled model displays a reduction in perception error difference between women and men, even though the reduction is limited. 
The reduction of the perception error between women and men suggests the existence of gender-preferred collaborations between gender groups.
Moreover, the Gender-shuffled model not only sensibly reduces the difference in the average perception error for women and men, but also achieves a nearly accurate perception on the fraction of women artists.
This strongly suggests that lowering an extent of imbalanced productivity and gendered preferences altogether boosts the representation of women artists, even considering the small representation of women artists in the company. 
%%%%

\section{Discussion}

Inequalities have been investigated for different occupations to capture the gender gap in salary and labor force composition~\cite{becker1985human}.
In this context, we find that the representation of women artists in ballet creations is about 18--22\%, which is lower than the reported 25\% for choreographic leads~\cite{dance_data}. 
These values are far below a gender-neutral ratio ($0.5$, see Table~\ref{tab:summary}) and are lower for highly central artists (in this study, TCA groups).
In general, we find that women artists are underrepresented in the overall collaboration network and all positions as leading artists, not only as choreographers. 
%%%%

In addition to the numerical imbalance, our results suggest that gendered collaboration structures could potentially aggravate gender imbalances in ballet creations. 
Crucial roles of individual's social network are associated with the access to information and professional opportunities in creative collaborations~\cite{luce1950connectivity, pan2012strength, abbasi2012egocentric, jamali2006different, uzzi_collaboration_2005, parish2018dynamics}.
Thus, our results increase the understanding of how gendered collaborations could impact artists' professional experiences and the social perceptions of women as a minority group. 
The comparison with null network models gives a hint that the observed gender imbalances in terms of central network positions could be explained by systematic inequalities in collaborative behaviors rather than by random factors.
%%%%%

The social network structures, such as network position and social prestige, play an important role shaping successful careers~\cite{fraiberger_quantifying_2018, janosov2020success}.
Some studies show that men and women utilize different social network structures and behavioral patterns that influence their placement in the job market~\cite{yang2019network, vasarhelyi2021gender, tata2008social}. 
Other studies show that the formation of a personal network and social behaviors over time are related to reinforced perception errors~\cite{jackson2014networks, ertan2019perception}. 
Taken together, these studies suggest the existence of a permanent feedback for the formation of social relationships, social perception errors, and collaboration patterns, which in return, can influence individual career decisions.
For women in ballet, a feedback based on a low representation within creative and men-dominated collaborations could negatively impact their decision to undertake a career as ballet creators or engaging in multiple collaborative projects.
A future study in this line could provide evidence on why women in ballet experience `glass barrier' in the field, and whether womens' network position facilitates their career success and impact in the long term.
%%%%

In addition, the collaboration structure can be crucial for teams~\cite{baugh1997effects} and individual performance~\cite{karakowsky2004perceptions} in terms of creativity and success~\cite{uzzi_collaboration_2005, genderdiverse_uzzi}. 
A study demonstrates that diversity can improve creative performance~\cite{hamilton2003team}, and emphasizes the participation of women in collaborative environments because they increase the social sensitivity of the group, making the team collectively more intelligent and proficient \cite{woolley2010evidence}. 
In view of this, new policies for more equal collaborations and a more inclusive environment for women as leading creators should be considered in ballet companies.
A more diverse inclusion would boost creative innovation and impact, which ultimately benefits the artistic community in general. 
%%%%

At an individual level, perception errors derived from the network structure in the workplace can affect career decisions.
In specific, our results reveal that most companies experience perception errors on the fraction of women artists participating at the company.
The constant underrepresentation of women could negavitely impact their visibility as a group, undermining the motivation of women artists to look for better professional opportunities. 
This interplay among working environment, perceived possibility of career development, and personal decisions could be a pivotal issue to alleviate the low representation of women in ballet and other industries where women are a minority.

Our measure of perception errors is a mathematical approach and can be improved, as there are multiple factors influencing the perception of a local network structure. 
That is to say, a local network can be described not only by its structure, but also to its embedded social mechanisms, like the strength of relationships formed over time, access to information, formal and informal norms~\cite{granovetter1973strength, granovetter1985economic, coleman1994foundations}, and individual cognitive processes and preferences~\cite{lazarsfeld1954friendship, pachur2013intuitive}.
In addition, ballet is strongly influenced by biological constraints, such as the physical demands of the art form, including strength, flexibility, and technical requirements. 
These constraints, combined with the distribution of labor in family responsibilities, may be stronger for women~\cite{becker1985human} and may contribute to fewer women to overcome social barriers in the workplace and hinder the professional development of women artists in ballet.
%%%%

Overall, our results help understand another dimension of gender inequalities in the ballet industry. 
Yet we are aware of the limitations of this work.
Our data depends on the archival of the selected ballet companies, which may not be sufficient to generalize the current results to entire ballet industry.
Moreover, artists may hold different types and duration of contracts within a company, which can result in variations in observed professional collaborations.
To overcome this, more comprehensive digitized data collections would be needed.
For instance, with the implementation of computational methods, such as deep learning and network science, it has been possible to objectively measure the impact of individual performance in creative domains~\cite{liu2018hot, liu2021understanding}, and similar methods can open the possibility for future research on the relationship between gender, network centrality, and actual ballet creators' impact in the field. 
%%%

In summary, our research highlights the low representation of women as ballet creators and sheds light on their peripheral network position and gendered collaboration preferences within the ballet industry. 
This investigation can be extended to explore dynamic network factors shaping gender imbalances to propose possible and more adequate interventions for the diversity, equity, and inclusion in cultural organizations. 
We hope that this work brings awareness on how social phenomena and inequalities in creative domains can be systematically studied with network science and data driven methods.

\subsection*{Data availability}

The data used in this study is available under reasonable request.

\subsection*{Abbreviations}
ABT, American Ballet Theatre; NBC, National Ballet of Canada; NYCB, New York City Ballet; ROH, Royal Ballet of the Royal Opera House; TCA, Top-Central Artists.

\subsection*{Availability of data and materials}
The data analyzed during the current study are available from the corresponding author on reasonable request.

\section*{Competing interests}
The authors declare that they have no competing interests.

\subsection*{Funding}
This work was supported by the National Research Foundation of Korea (NRF) grant funded by the Korea government (MSIT) (No.NRF-2022R1C1C1005856), the National Agency of Investigation and Development, ANID, through the grant FONDECYT No. 11190096, the KENTECH Research Grant(KRG 2021-01-003), and the Pukyong National University Research Fund in 2022(202203530001).

\subsection*{Author's contributions}

All authors contributed to the research design and writing of the paper.
YHG contributed with art-specific knowledge, constructed the data and networks, developed and performed the models, analyzed the data, and performed data visualizations; EL was mainly responsible for the measurement of perception errors and homophily; and HK contributed to data construction and network analysis.
EL and HK supervised the research. 
All authors discussed the results and contributed to writing the manuscript.
All authors read and approved the final manuscript.

\subsection*{Acknowledgements}

YHG acknowledges the Centro de Investigación en Complejidad at Universidad del Desarrollo, Chile, for the financial support to conduct this research.
%%%%

\printbibliography[heading=subbibintoc]

@article{becker1985human,
	author = {Becker, Gary S},
	journal = {Journal of labor economics},
	number = {1, Part 2},
	pages = {S33--S58},
	publisher = {University of Chicago Press},
	title = {Human capital, effort, and the sexual division of labor},
	volume = {3},
	year = {1985},
    url = {https://www.jstor.org/stable/2534997}
	}

@book{atkinson2015inequality,
	author = {Atkinson, Anthony B},
    ISBN = {9780674979789},
	publisher = {Harvard University Press},
	title = {Inequality: What can be done?},
	year = {2015}}

@article{piketty2015capital,
	author = {Piketty, Thomas},
	journal = {American Economic Review},
	number = {5},
	pages = {48--53},
	title = {About capital in the twenty-first century},
	volume = {105},
	year = {2015},
    doi={10.4159/9780674982918}}

@article{topaz2019diversity,
	author = {Topaz, Chad M and Klingenberg, Bernhard and Turek, Daniel and Heggeseth, Brianna and Harris, Pamela E and Blackwood, Julie C and Chavoya, C Ondine and Nelson, Steven and Murphy, Kevin M},
	journal = {PloS one},
	number = {3},
	pages = {e0212852},
	publisher = {Public Library of Science},
	title = {Diversity of artists in major US museums},
	volume = {14},
	year = {2019},
    doi = {10.1371/journal.pone.0212852}}

@article{lindemann2016asymmetrical,
	author = {Lindemann, Danielle J and Rush, Carly A and Tepper, Steven J},
	journal = {Social Currents},
	number = {4},
	pages = {332--348},
	publisher = {Sage Publications Sage CA: Los Angeles, CA},
	title = {An asymmetrical portrait: Exploring gendered income inequality in the arts},
	volume = {3},
	year = {2016},
    doi = {10.1177/2329496516636399}}

@article{daly1987balanchine,
	author = {Daly, Ann},
	journal = {The drama review: TDR},
	number = {1},
	pages = {8--21},
	publisher = {JSTOR},
	title = {The Balanchine woman: Of hummingbirds and channel swimmers},
	volume = {31},
	year = {1987},
    doi = {10.2307/1145763},
    url = {https://www.jstor.org/stable/1145763}}

@book{homans2010apollo,
	author = {Homans, Jennifer},
	edition = {1st edition},
	publisher = {Random House Publishing Group},
	title = {Apollo's angels: A history of ballet},
	year = {2010},
    ISBN = {9780812968743}}

@article{defrank2016slow,
	author = {DeFrank-Cole, Lisa and Nicholson, Renee K},
	journal = {Leadership and the Humanities},
	number = {2},
	pages = {73--91},
	publisher = {Edward Elgar Publishing Ltd},
	title = {The slow-changing face of leadership in ballet: an interdisciplinary approach to analysing women's roles},
	volume = {4},
	year = {2016},
    doi = {10.4337/lath.2016.02.01}}

@misc{gendergapballet,
	author = {Kim Elsesser},
	month = {09},
	title = {A Gender Gap In Ballet, Seriously?},
	url = {https://www.forbes.com/sites/kimelsesser/2019/09/12/a-gender-gap-in-ballet-seriously/},
	year = {2019}}

@misc{balletworld,
	author = {Elizabeth Yntema},
	month = {08},
	title = {The ballet world is still male-dominated, research shows},
	url = {https://www.womensmediacenter.com/news-features/the-ballet-world-is-still-male-dominated-research-shows},
	year = {2019}}

@misc{dance_data,
	author = {},
	lastchecked = {January 2021},
	title = {Dance Data Project},
	url = {www.dancedataproject.com},
	year = {2015}}

@misc{bureau_data,
	author = {},
	lastchecked = {July 2021},
    title = {United States Census Bureau}, 
	url = {https://www.census.gov/data.html},
	year = {2021}
 }

@article{cotter2001glass,
	author = {Cotter, David A and Hermsen, Joan M and Ovadia, Seth and Vanneman, Reeve},
	journal = {Social forces},
	number = {2},
	pages = {655--681},
	publisher = {Oxford University Press},
	title = {The glass ceiling effect},
	volume = {80},
	year = {2001},
    doi = {10.1353/sof.2001.0091}}

@article{maslov2002specificity,
	author = {Maslov, Sergei and Sneppen, Kim},
	journal = {Science},
	number = {5569},
	pages = {910--913},
	publisher = {American Association for the Advancement of Science},
	title = {Specificity and stability in topology of protein networks},
	volume = {296},
	year = {2002},
    doi = {10.1126/science.1065103}}

@book{wasserman1994,
	author = {Wasserman, Stanley and Faust, Katherine},
	publisher = {Cambridge University Press},
	series = {Structural Analysis in the Social Sciences},
	title = {Social network analysis: Methods and applications},
	year = {1994},
    doi = {10.1017/CBO9780511815478}}

@article{mcpherson2001birds,
	author = {McPherson, Miller and Smith-Lovin, Lynn and Cook, James M},
	journal = {Annual Review of Sociology},
	number = {1},
	pages = {415--444},
	publisher = {Annual Reviews 4139 El Camino Way, PO Box 10139, Palo Alto, CA 94303-0139, USA},
	title = {Birds of a feather: Homophily in social networks},
	volume = {27},
	year = {2001},
    doi = {10.1146/annurev.soc.27.1.415}}

@article{liu2021understanding,
	author = {Liu, Lu and Dehmamy, Nima and Chown, Jillian and Giles, C Lee and Wang, Dashun},
	journal = {Nature communications},
	number = {1},
	pages = {1--10},
	publisher = {Nature Publishing Group},
	title = {Understanding the onset of hot streaks across artistic, cultural, and scientific careers},
	volume = {12},
	year = {2021},
    doi = {10.1038/s41467-021-25477-8}}

@article{liu2018hot,
	author = {Liu, Lu and Wang, Yang and Sinatra, Roberta and Giles, C Lee and Song, Chaoming and Wang, Dashun},
	journal = {Nature},
	number = {7714},
	pages = {396--399},
	publisher = {Nature Publishing Group},
	title = {Hot streaks in artistic, cultural, and scientific careers},
	volume = {559},
	year = {2018},
    doi = {10.1038/s41586-018-0315-8}}

@article{genderdiverse_uzzi,
	author = {Yang Yang and Tanya Y. Tian and Teresa K. Woodruff and Benjamin F. Jones and Brian Uzzi},
	doi = {10.1073/pnas.2200841119},
	journal = {Proceedings of the National Academy of Sciences},
	number = {36},
	pages = {e2200841119},
	title = {Gender-diverse teams produce more novel and higher-impact scientific ideas},
	url = {https://www.pnas.org/doi/abs/10.1073/pnas.2200841119},
	volume = {119},
	year = {2022}}

@article{pachur2013intuitive,
	author = {Pachur, Thorsten and Hertwig, Ralph and Rieskamp, J{\"o}rg},
	journal = {Journal of Experimental Social Psychology},
	number = {6},
	pages = {1059--1077},
	publisher = {Elsevier},
	title = {Intuitive judgments of social statistics: How exhaustive does sampling need to be?},
	volume = {49},
	year = {2013},
    doi = {10.1016/j.jesp.2013.07.004}}

@article{lazarsfeld1954friendship,
	author = {Lazarsfeld, Paul F and Merton, Robert K},
	journal = {Freedom and Control in Modern Society},
	number = {1},
	pages = {18--66},
	publisher = {New York, Van Nostrand},
	title = {Friendship as a social process: A substantive and methodological analysis},
	volume = {18},
	year = {1954}}

@book{coleman1994foundations,
	author = {Coleman, James S},
    ISBN = {9780674312265},
	publisher = {Harvard University Press},
	title = {Foundations of Social Theory},
	year = {1994},
    doi = {10.2307/2579680},
    url ={https://www.jstor.org/stable/2579680}}

@article{granovetter1985economic,
	author = {Granovetter, Mark},
	journal = {American Journal of Sociology},
	number = {3},
	pages = {481--510},
	publisher = {University of Chicago Press},
	title = {Economic action and social structure: The problem of embeddedness},
	volume = {91},
	year = {1985},
    url = {https://www.jstor.org/stable/2780199}}

@article{granovetter1973strength,
	author = {Granovetter, Mark S},
	journal = {American Journal of Sociology},
	number = {6},
	pages = {1360--1380},
	publisher = {University of Chicago Press},
	title = {The strength of weak ties},
	volume = {78},
	year = {1973},
    url = {https://www.jstor.org/stable/2776392}}

@article{woolley2010evidence,
	author = {Woolley, Anita Williams and Chabris, Christopher F and Pentland, Alex and Hashmi, Nada and Malone, Thomas W},
	journal = {Science},
	number = {6004},
	pages = {686--688},
	publisher = {American Association for the Advancement of Science},
	title = {Evidence for a collective intelligence factor in the performance of human groups},
	volume = {330},
	year = {2010},
    doi = {10.1126/science.1193147}}

@article{hamilton2003team,
	author = {Hamilton, Barton H and Nickerson, Jack A and Owan, Hideo},
	journal = {Journal of Political Economy},
	number = {3},
	pages = {465--497},
	publisher = {The University of Chicago Press},
	title = {Team incentives and worker heterogeneity: An empirical analysis of the impact of teams on productivity and participation},
	volume = {111},
	year = {2003},
    url = {https://www.jstor.org/stable/10.1086/374182}}

@article{baugh1997effects,
	author = {Baugh, S Gayle and Graen, George B},
	journal = {Group \& Organization Management},
	number = {3},
	pages = {366--383},
	publisher = {Sage Publications Sage CA: Thousand Oaks, CA},
	title = {Effects of team gender and racial composition on perceptions of team performance in cross-functional teams},
	volume = {22},
	year = {1997},
	doi = {10.1177/1059601197223004}}

@article{karakowsky2004perceptions,
	author = {Karakowsky, Leonard and McBey, Kenneth and Chuang, You-Ta},
    ISSN = {0268-3946},
	journal = {Journal of Managerial Psychology},
	publisher = {Emerald Group Publishing Limited},
	title = {Perceptions of team performance: The impact of group composition and task-based cues},
	year = {2004},
    doi = {10.1108/02683940410543597}}

@article{ertan2019perception,
	author = {Ertan, G{\"u}ne{\c{s}} and Siciliano, Michael D and Yenig{\"u}n, Deniz},
	journal = {PloS One},
	number = {6},
	pages = {e0218607},
	publisher = {Public Library of Science San Francisco, CA USA},
	title = {Perception accuracy, biases and path dependency in longitudinal social networks},
	volume = {14},
	year = {2019},
    doi = {10.1371/journal.pone.0218607}}

@article{jackson2014networks,
	author = {Jackson, Matthew O},
	journal = {Journal of Economic Perspectives},
	number = {4},
	pages = {3--22},
	title = {Networks in the understanding of economic behaviors},
	volume = {28},
	year = {2014},
    doi = {10.1257/jep.28.4.3}}

@article{tata2008social,
	author = {Tata, Jasmine and Prasad, Sameer},
	journal = {International Journal of Entrepreneurship and Small Business},
	number = {3-4},
	pages = {373--388},
	publisher = {Inderscience Publishers},
	title = {Social capital, collaborative exchange and microenterprise performance: The role of gender},
	volume = {5},
	year = {2008},
    doi = {10.1504/IJESB.2008.01731}}

@article{vasarhelyi2021gender,
	author = {Vasarhelyi, Orsolya and Vedres, Balazs},
	journal = {arXiv preprint arXiv:2103.01093},
	title = {Gender Typicality of Behavior Predicts Success on Creative Platforms},
	year = {2021},
    doi = {10.48550/arXiv.2103.01093}}

@article{janosov2020success,
	author = {Janosov, Mil{\'a}n and Battiston, Federico and Sinatra, Roberta},
	journal = {EPJ Data Science},
	number = {1},
	pages = {9},
	publisher = {Springer Berlin Heidelberg},
	title = {Success and luck in creative careers},
	volume = {9},
	year = {2020},
    doi = {10.1140/epjds/s13688-020-00227-w}}

@article{fraiberger_quantifying_2018,
	author = {Fraiberger, Samuel P. and Sinatra, Roberta and Resch, Magnus and Riedl, Christoph and Barab{\'a}si, Albert-L{\'a}szl{\'o}},
	doi = {10.1126/science.aau7224},
	issn = {0036-8075, 1095-9203},
	journal = {Science},
	language = {en},
	month = {11},
	number = {6416},
	pages = {825--829},
	title = {Quantifying reputation and success in art},
	url = {http://www.sciencemag.org/lookup/doi/10.1126/science.aau7224},
	urldate = {2019-01-28},
	volume = {362},
	year = {2018}}

@inproceedings{jamali2006different,
	author = {Jamali, Mohsen and Abolhassani, Hassan},
	booktitle = {2006 IEEE/WIC/ACM International Conference on Web Intelligence (WI 2006 Main Conference Proceedings)(WI'06)},
	organization = {IEEE},
	pages = {66--72},
	title = {Different aspects of social network analysis},
	year = {2006},
    doi = {10.1109/WI.2006.61}}

@article{luce1950connectivity,
	author = {Luce, R Duncan},
	journal = {Psychometrika},
	number = {2},
	pages = {169--190},
	publisher = {Springer},
	title = {Connectivity and generalized cliques in sociometric group structure},
	volume = {15},
	year = {1950},
    doi = {10.1007/BF02289199}}

@article{bonacich2007some,
	author = {Bonacich, Phillip},
	journal = {Social Networks},
	number = {4},
	pages = {555--564},
	publisher = {Elsevier},
	title = {Some unique properties of eigenvector centrality},
	volume = {29},
	year = {2007},
    doi = {10.1016/j.socnet.2007.04.002}}

@article{freeman1978centrality,
	author = {Freeman, Linton C},
	journal = {Social Networks},
	number = {3},
	pages = {215--239},
	publisher = {North-Holland},
	title = {Centrality in Social Networks: I. Conceptual Clarification},
	volume = {1},
	year = {1978},
    doi = {10.1016/0378-8733(78)90021-7}}

@article{boldi2014axioms,
	author = {Boldi, Paolo and Vigna, Sebastiano},
	journal = {Internet Mathematics},
	number = {3-4},
	pages = {222--262},
	publisher = {Taylor \& Francis},
	title = {Axioms for centrality},
	volume = {10},
	year = {2014},
    doi = {10.1080/15427951.2013.865686}}

@article{freeman1977set,
	author = {Freeman, Linton C},
	journal = {Sociometry},
	pages = {35--41},
	publisher = {JSTOR},
	title = {A set of measures of centrality based on betweenness},
	year = {1977},
    doi = {10.2307/3033543},
    url = {https://www.jstor.org/stable/3033543}}

@article{blevins2015jane,
	author = {Blevins, Cameron and Mullen, Lincoln},
	journal = {DHQ: Digital Humanities Quarterly},
	number = {3},
	title = {Jane, John... Leslie? A Historical Method for Algorithmic Gender Prediction.},
	volume = {9},
	year = {2015},
    note = {R package version 0.6.0},
    url = {https://github.com/lmullen/gender}}

@inproceedings{karimi2016inferring,
	author = {Karimi, Fariba and Wagner, Claudia and Lemmerich, Florian and Jadidi, Mohsen and Strohmaier, Markus},
	booktitle = {Proceedings of the 25th International conference companion on World Wide Web},
	pages = {53--54},
	title = {Inferring gender from names on the web: A comparative evaluation of gender detection methods},
	year = {2016},
    doi = {10.1145/2872518.2889385}}

@article{humphries2008network,
	author = {Humphries, Mark D and Gurney, Kevin},
	journal = {PloS One},
	number = {4},
	pages = {e0002051},
	publisher = {Public Library of Science},
	title = {Network `small-world-ness': a quantitative method for determining canonical network equivalence}, 
	volume = {3},
	year = {2008},
    doi = {10.1371/journal.pone.0002051}}

@article{chen2008average,
	author = {Chen, Fei and Chen, Zengqiang and Wang, Xiufeng and Yuan, Zhuzhi},
	journal = {Communications in Nonlinear Science and numerical simulation},
	number = {7},
	pages = {1405--1410},
	publisher = {Elsevier},
	title = {The average path length of scale free networks},
	volume = {13},
	year = {2008},
    doi = {10.1016/j.cnsns.2006.12.003}}

@article{watts1998collective,
	author = {Watts, Duncan J and Strogatz, Steven H},
	journal = {nature},
	number = {6684},
	pages = {440--442},
	publisher = {Nature Publishing Group},
	title = {Collective dynamics of `small-world' networks},
	volume = {393},
	year = {1998},
    doi = {10.1038/30918}}

@misc{van2018robotic,
	author = {Van der Aalst, Wil MP and Bichler, Martin and Heinzl, Armin},
	publisher = {Springer},
	title = {Robotic Process Automation},
	year = {2018},
    doi = {10.1007/s12599-018-0542-4}}

@misc{ROH,
	author = {},
	lastchecked = {April 2021},
	title = {Royal Opera House Collections},
	url = {http://rohcollections.org.uk}}

@misc{NBC,
	author = {},
	lastchecked = {April 2021},
	title = {The National Ballet of Canada Archives, Repertoire List},
	url = {https://national.ballet.ca/Tickets/Archives/Repertoire-List}}

@misc{NYCB,
	author = {},
	lastchecked = {April 2021},
	title = {New York City Ballet, The Repertory},
	url = {https://www.nycballet.com/discover/ballet-repertory}}

@misc{ABT,
    author = "",
    title = "American Ballet Theatre, Ballet Archive",
    url  = "https://www.abt.org/explore/learn/repertory-archive/ballets/",
    addendum = "(accessed: 04.09.2021)"
}

@article{lee2019homophily,
	author = {Lee, Eun and Karimi, Fariba and Wagner, Claudia and Jo, Hang-Hyun and Strohmaier, Markus and Galesic, Mirta},
	journal = {Nature human behaviour},
	number = {10},
	pages = {1078--1087},
	publisher = {Nature Publishing Group},
	title = {Homophily and minority-group size explain perception biases in social networks}, 
	volume = {3},
	year = {2019},
    doi = {10.1038/s41562-019-0677-4}}

@article{lerman2016majority,
	author = {Lerman, Kristina and Yan, Xiaoran and Wu, Xin-Zeng},
	journal = {PLoS One},
	number = {2},
	pages = {e0147617},
	title = {The `Majority Illusion' in Social Networks}, 
	volume = {11},
	year = {2016},
    doi = {10.1371/journal.pone.0147617}}

@article{festinger1954theory,
	author = {Festinger, Leon},
	journal = {Human Relations},
	number = {2},
	pages = {117--140},
	publisher = {Sage Publications Sage CA: Thousand Oaks, CA},
	title = {A theory of social comparison processes},
	volume = {7},
	year = {1954},
    doi = {10.1177/001872675400700202}}

@article{karimi2018homophily,
	author = {Karimi, Fariba and G{\'e}nois, Mathieu and Wagner, Claudia and Singer, Philipp and Strohmaier, Markus},
	journal = {Scientific Reports},
	number = {1},
	pages = {1--12},
	publisher = {Nature Publishing Group},
	title = {Homophily influences ranking of minorities in social networks},
	volume = {8},
	year = {2018},
    doi = {10.1038/s41598-018-29405-7}}

@article{yang2019network,
	author = {Yang, Yang and Chawla, Nitesh V and Uzzi, Brian},
	journal = {Proceedings of the National Academy of Sciences},
	number = {6},
	pages = {2033--2038},
	publisher = {National Acad Sciences},
	title = {A network's gender composition and communication pattern predict women's leadership success},
	volume = {116},
	year = {2019},
    doi = {10.1073/pnas.1721438116}}

@article{parish2018dynamics,
	author = {Parish, Austin J and Boyack, Kevin W and Ioannidis, John PA},
	journal = {PloS One},
	number = {1},
	pages = {e0189742},
	publisher = {Public Library of Science San Francisco, CA USA},
	title = {Dynamics of co-authorship and productivity across different fields of scientific research},
	volume = {13},
	year = {2018},
    doi = {10.1371/journal.pone.0189742}}

@article{juhasz2020brokering,
	author = {Juh{\'a}sz, S{\'a}ndor and T{\'o}th, Gerg{\H{o}} and Lengyel, Bal{\'a}zs},
	journal = {PloS one},
	number = {2},
	pages = {e0229436},
	publisher = {Public Library of Science San Francisco, CA USA},
	title = {Brokering the core and the periphery: Creative success and collaboration networks in the film industry},
	volume = {15},
	year = {2020},
    doi = {10.1371/journal.pone.0229436}}

@article{abbasi2012egocentric,
	author = {Abbasi, Alireza and Chung, Kon Shing Kenneth and Hossain, Liaquat},
	journal = {Information Processing \& Management},
	number = {4},
	pages = {671--679},
	publisher = {Elsevier},
	title = {Egocentric analysis of co-authorship network structure, position and performance},
	volume = {48},
	year = {2012},
    doi = {10.1016/j.ipm.2011.09.001}}

@article{uzzi_collaboration_2005,
	author = {Uzzi, Brian and Spiro, Jarrett},
	doi = {10.1086/432782},
	issn = {0002-9602, 1537-5390},
	journal = {American Journal of Sociology},
	language = {en},
	month = sep,
	number = {2},
	pages = {447--504},
	shorttitle = {Collaboration and {Creativity}},
	title = {Collaboration and Creativity: The Small World Problem},
	url = {http://www.journals.uchicago.edu/doi/10.1086/432782},
	urldate = {2019-04-02},
	volume = {111},
	year = {2005}}

@article{pan2012strength,
	author = {Pan, Raj Kumar and Saram{\"a}ki, Jari},
	journal = {EPL (Europhysics Letters)},
	number = {1},
	pages = {18007},
	publisher = {IOP Publishing},
	title = {The strength of strong ties in scientific collaboration networks},
	volume = {97},
	year = {2012},
    doi = {10.1209/0295-5075/97/18007}}

@article{williams1992glass,
	author = {Williams, Christine L},
	journal = {Social Problems},
	pages = {253--267},
	title = {The glass escalator: Hidden advantages for men in non traditional occupations},
	volume = {39},
	year = {1992},
    doi = {10.2307/3096961},
    url = {https://www.jstor.org/stable/3096961}}

\clearpage

\begin{center}
\Large{Supplementary Information for\\[0.8cm]
\Large\textbf{Structural Gender Imbalances\\[0.2cm] in Ballet Collaboration Networks}}\\[0.2cm]

{\small Yessica Herrera-Guzm\'{a}n, Eun Lee, Heetae Kim} \\[0.2cm]

{\small *Corresponding Author. Email: \href{hkim@kentech.ac.kr}{hkim@kentech.ac.kr}}
\end{center}

% \pagenumbering{arabic}

%%%%%%%%%%%%%%%%%%%%%%%%%%%%%%%%%%%%%%%%
\renewcommand{\thefigure}{S\arabic{figure}}
\setcounter{figure}{0}
\renewcommand{\theequation}{S\arabic{equation}}
\setcounter{equation}{0}
\renewcommand{\thesection}{S\arabic{section}}
\setcounter{section}{0}
\renewcommand{\thetable}{S\arabic{table}}
\setcounter{table}{0}
\newcounter{SIfig}
\renewcommand{\theSIfig}{S\arabic{SIfig}}
%%%%%%%%%%%%%%%%%%%%%%%%%

\bigskip

\vspace{30pt}

\setstretch{1.5}
\noindent
\textbf{This file includes:}

Supplementary Text

Figures S1 to S7

% \clearpage

\section{\large Ballet collaborations}
\label{section:collaborations}

We define a \emph{ballet collaboration} as the creative effort between choreographers, composers, costume designers, lighting designers, and other artists listed by each company, for the creation of a ballet work.
In practice, other types of ballet collaborations are possible, such as those that put the ballet work on stage and are involved in the development of a production (e.g. production managers, technicians, theatre staff). 
Also, some ballet creations require the direct collaboration of the choreographers with the ballet dancers. 
Due to limited access to company data, we only consider the artistic roles as defined in a ballet collaboration. 
%%

% Add other company constraints. 

% Add the representation by gender in each role, and explain why we are not considering that for this research. 

%%%%%%%%%%%%%%%%%%%%%%%%%%%%%%%%%%%%%%%%%%%%%%%%%%%%%
\begin{figure}[!ht]
    \centering
    \caption{\textbf{Team composition.} Panel \textbf{a} shows the probability of a team size in each ballet company. Panel \textbf{b} shows the probability of a gender ratio in a team.}
    \includegraphics[scale=0.6]{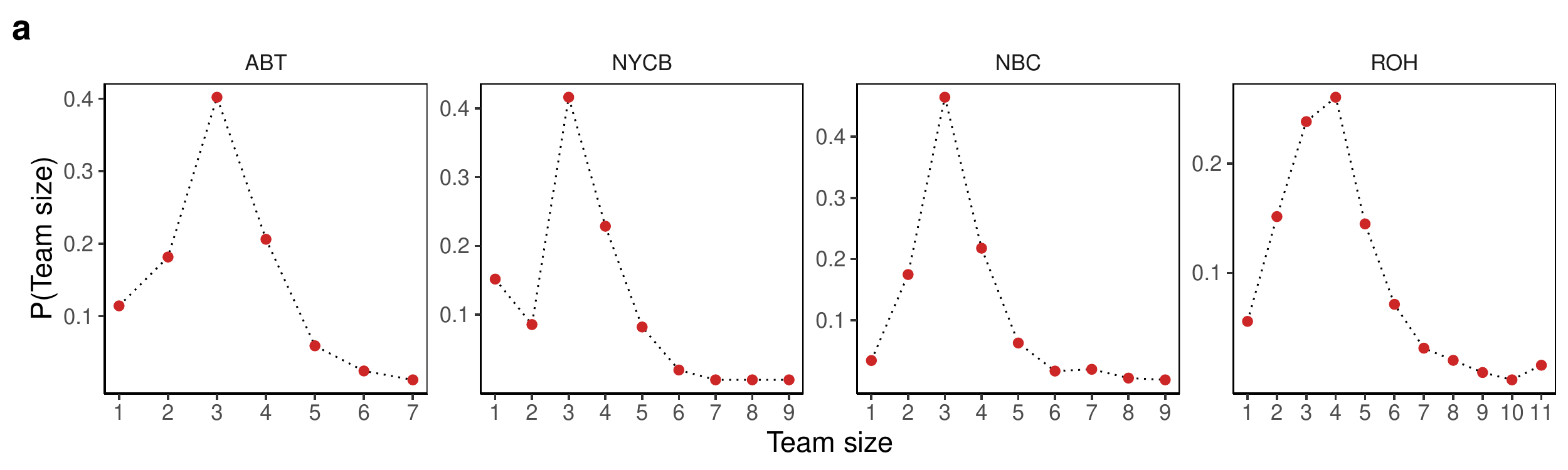}\\
    \includegraphics[scale=0.6]{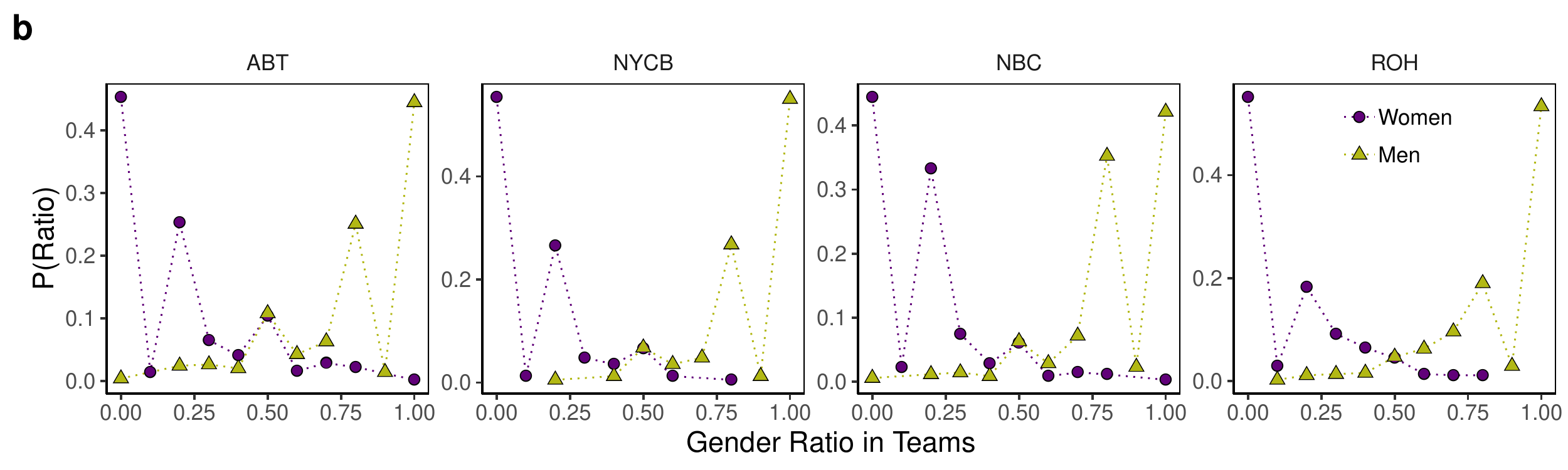}\\
    \refstepcounter{SIfig}\label{Sfig_supp:teamsize}
\end{figure}
%%%%%%%%%%%%%%%%%%%%%%%%%%%%%%%%%%%%%%%%%%%%%%%%%%%%%

%%%%%%%%%%%%%%%%%%%%%%%%%%%%%%%%%%%%%%%%
\begin{figure}[!ht]
\centering
    \caption{\textbf{Representation of women in ballet collaborations by artistic roles.}  
    Fraction of women written inside purple bar. 
    Group size indicated by bar-width with total count of artists written on top of bar. 
    Figures show women's representation by artistic roles for \textbf{a} ABT, \textbf{b} NYCB, \textbf{c} NBC, and \textbf{d} ROH.
    While women face less representation in the Composers group, their representation is consistently low across artistic roles, considering group size.}
    \includegraphics[scale=0.36]{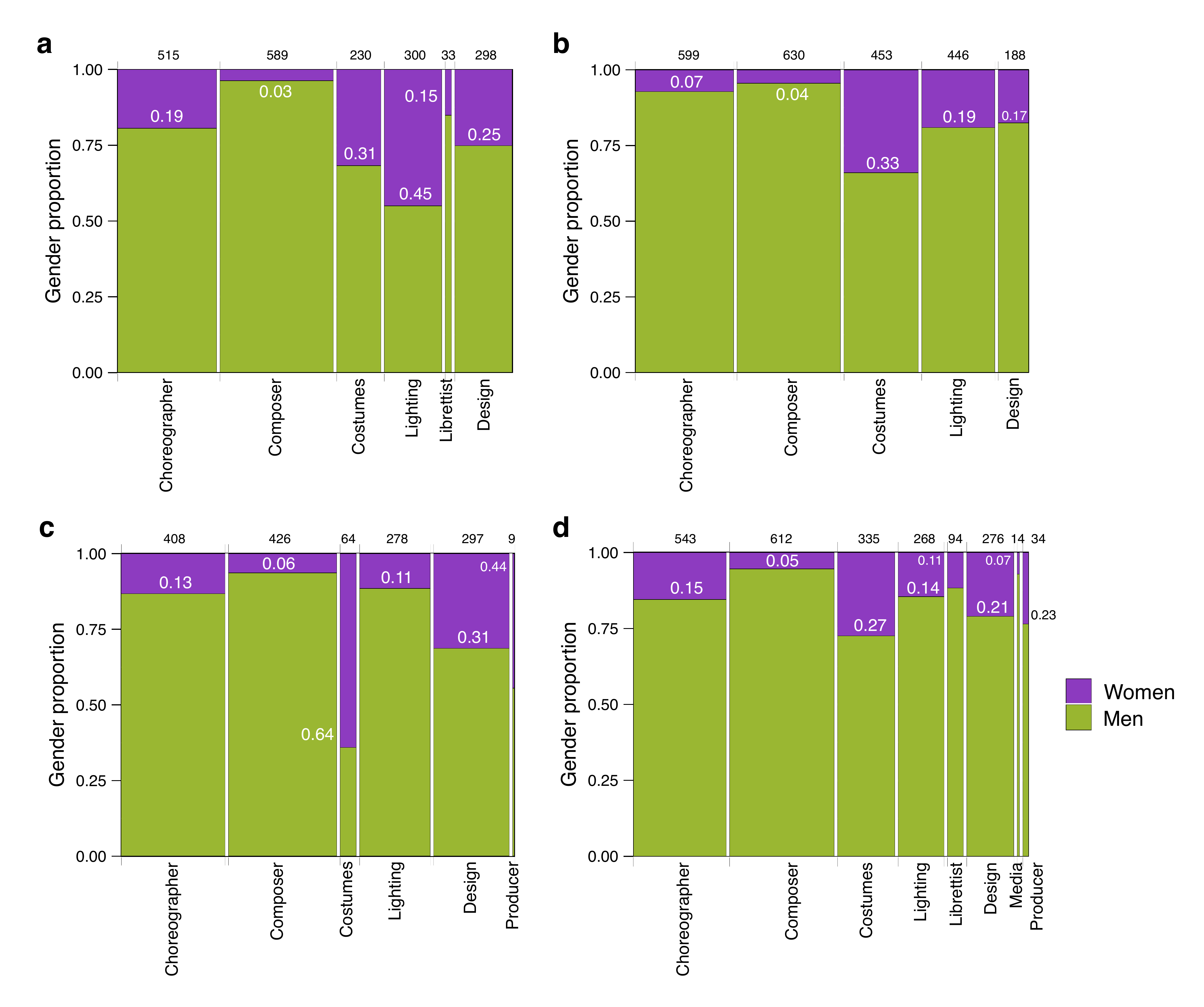}\\
\refstepcounter{SIfig}\label{Sfig_supp:women_artist}
\end{figure}
%%%%%%%%%%%%%%%%%%%%%%%%%%%%%%%%%%%%%%%%%%

%%%%%%%%%%%%%%%%%%%%%%%%%%%%%%%%%%%%%%%%%%%%%%%%%%%%%
\begin{figure}[!ht]
\centering
\caption{\textbf{Collaboration patterns by gender.} Purple/green and dots/triangles represent women/men. Panel \textbf{a} shows the fraction of teams by the number of same gender artists in their composition. Panel \textbf{b} shows the number of ballet creations and the count of artists by gender (artists’ productivity).}
\includegraphics[width =0.95\textwidth]{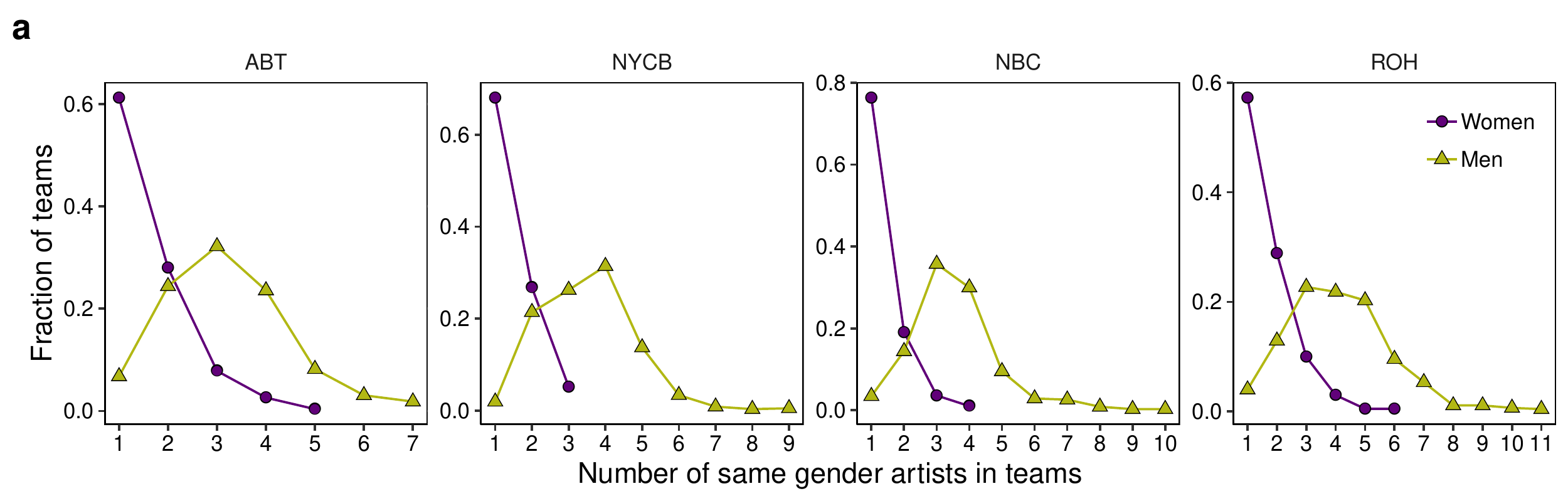}\\
\includegraphics[width =0.95\textwidth]{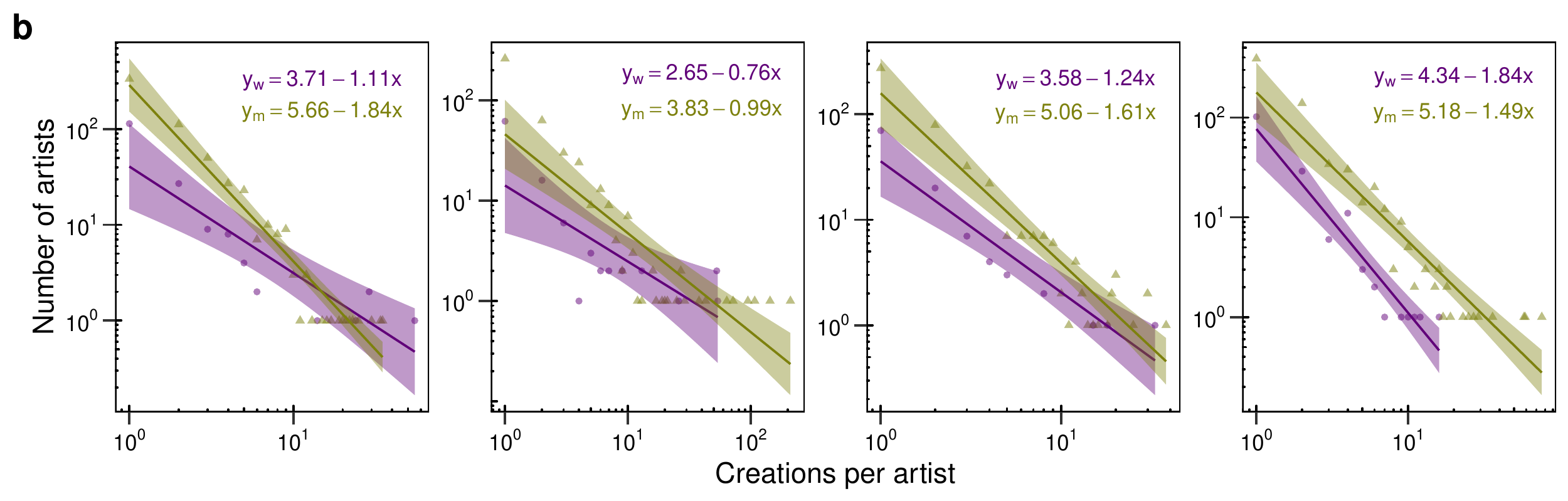}\\
\refstepcounter{SIfig}\label{Sfig_supp:collaborations}
\end{figure}
%%%%%%%%%%%%%%%%%%%%%%%%%%%%%%%%%%%%%%%%%%%%%%%%%%%%%

\clearpage

\section{\large Centrality differences by gender and null models analyses}
\label{section:networks}

%%%%%%%%%%%%%%%%%%%%%%%%%%%%%%%%%%%%%%%%%%%%%%%%%%%%%
\begin{figure}[ht!]
\centering
\caption{\textbf{Distribution of artists in the collaboration networks from ABT, NYCB, and NBC.}}
\begin{enumerate}[label=(\alph*)]
\centering
\item \textbf{Degree distribution.}
\end{enumerate}
\vspace{-6pt}
\begin{tabular}{ccc}
\setlength{\tabcolsep}{-20pt}
\includegraphics[width=0.28\textwidth]{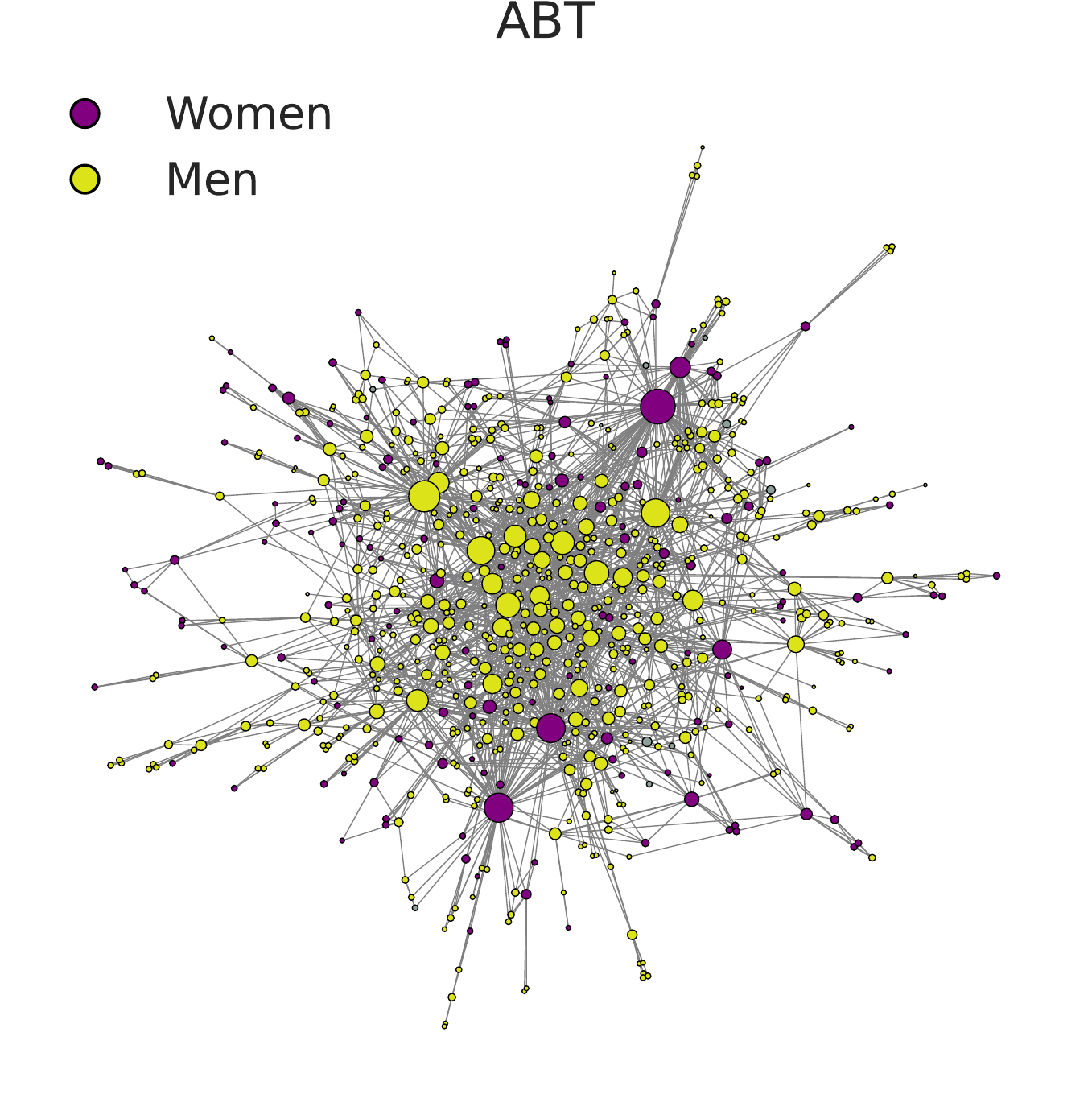} & \includegraphics[width=0.28\textwidth]{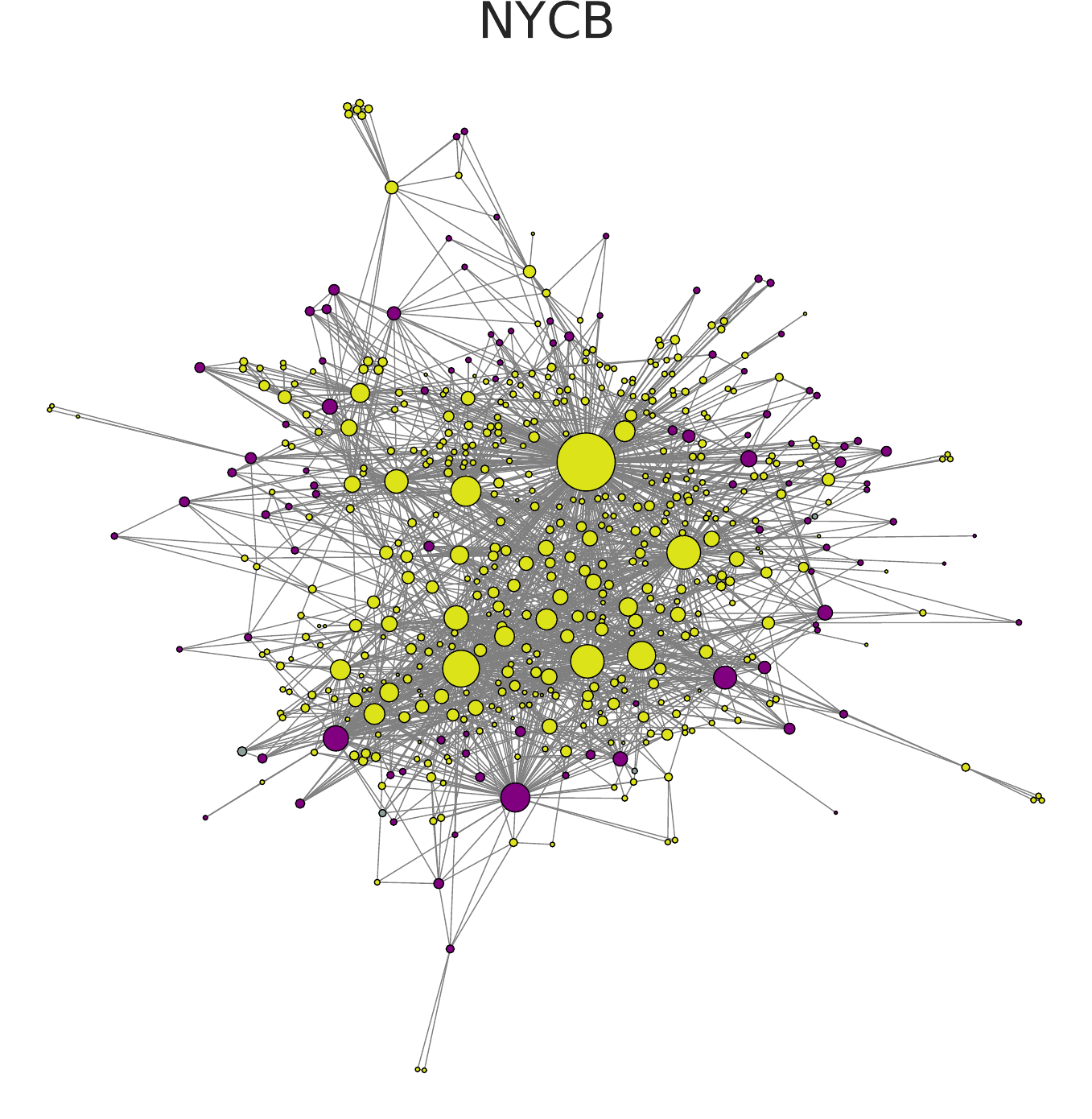} & \includegraphics[width=0.28\textwidth]{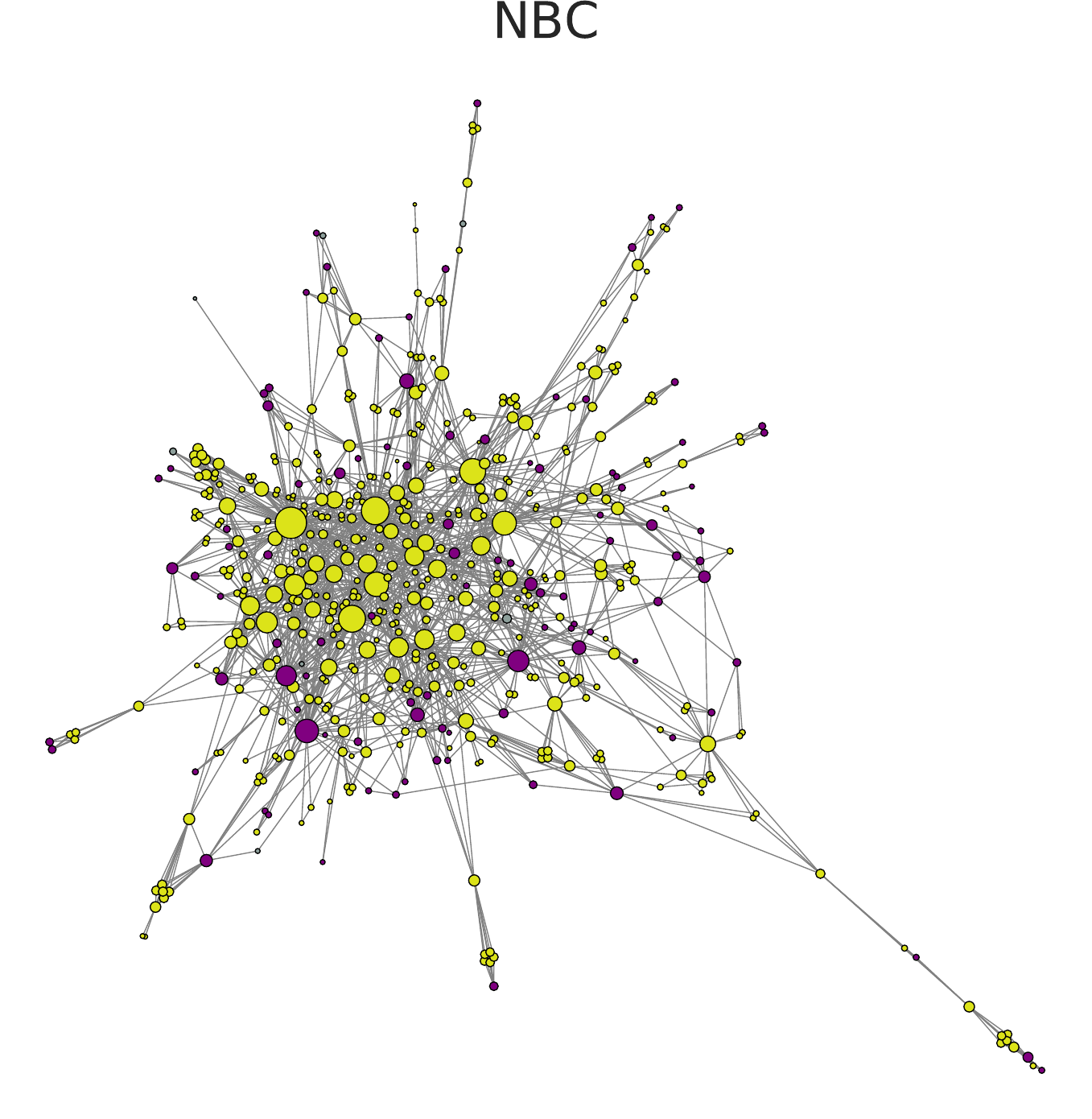} \\
\end{tabular}

\begin{enumerate}[label=(\alph*), start=2]
\centering
\item \textbf{Network position of man-man collaborations.}
\end{enumerate}
\vspace{-6pt}
\begin{tabular}{ccc}
\includegraphics[width=0.28\textwidth]{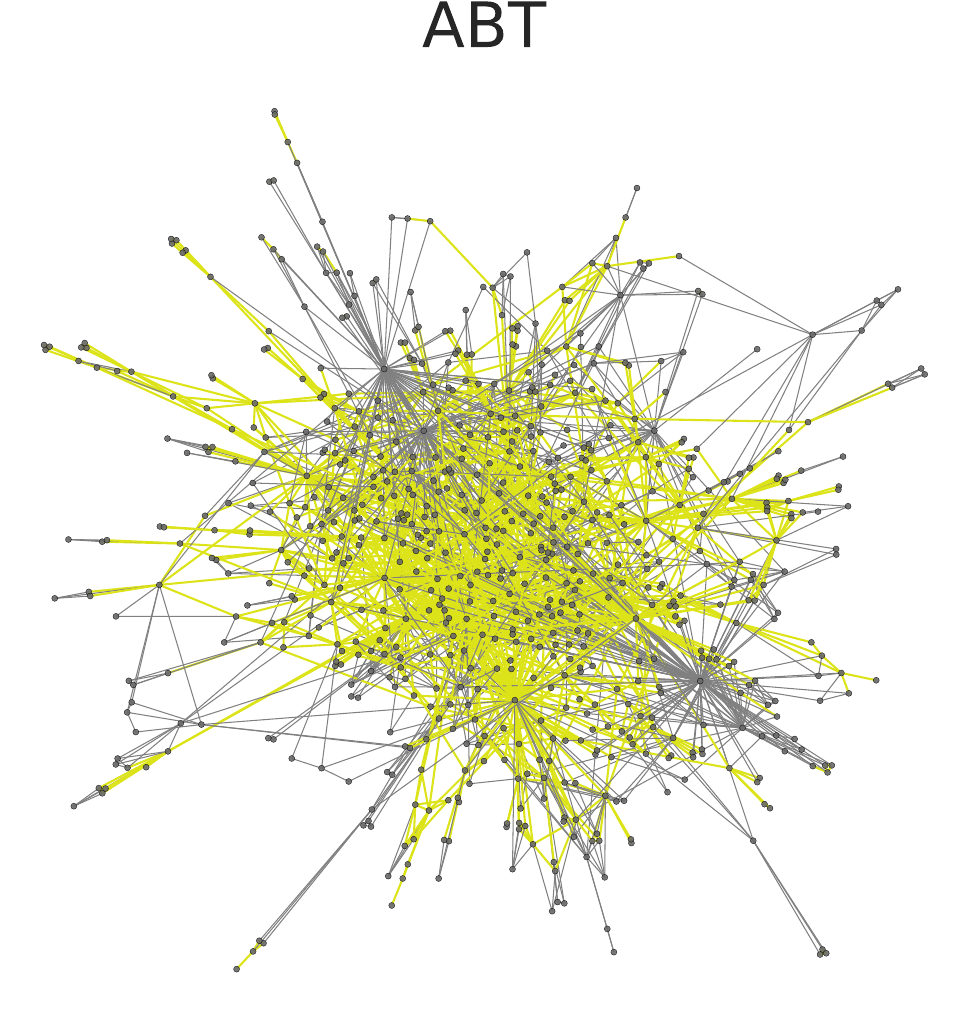} & \includegraphics[width=0.28\textwidth]{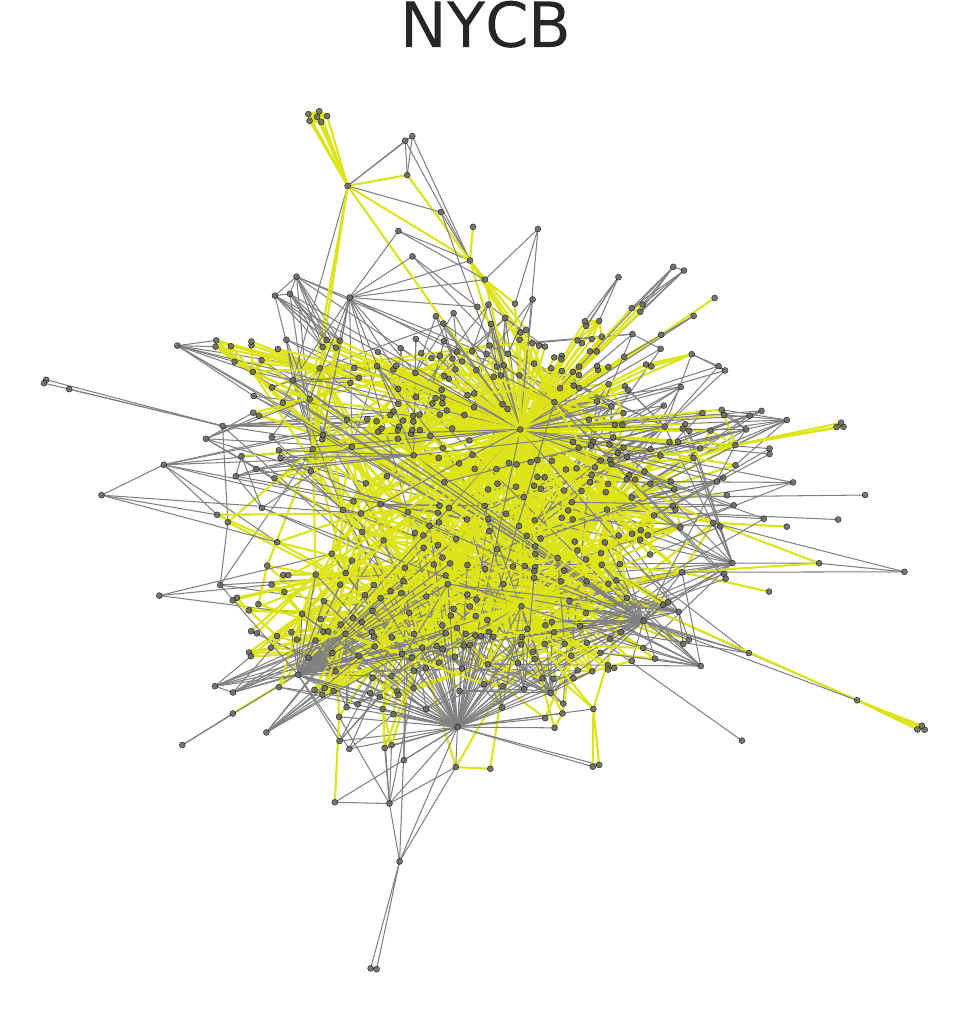} & \includegraphics[width=0.28\textwidth]{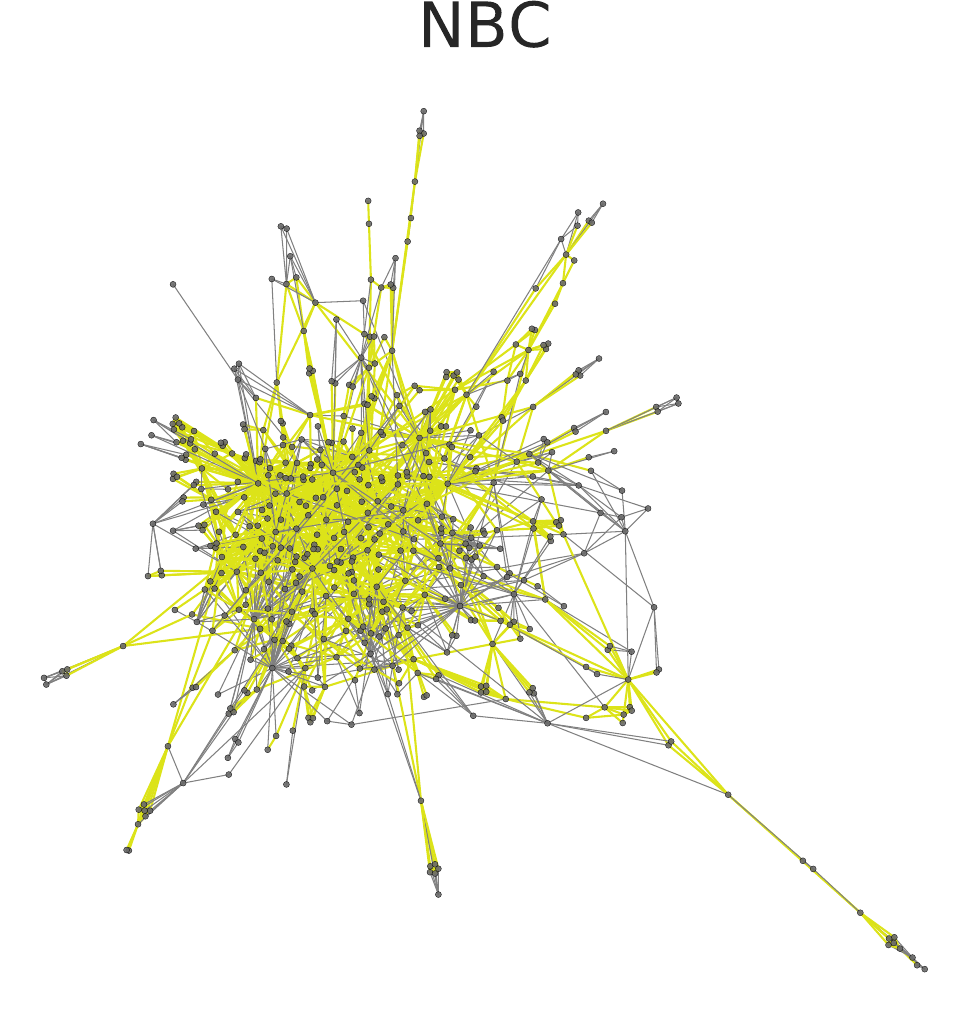} \\
\end{tabular}

\vspace{-12pt}
\begin{enumerate}[label=(\alph*), start=3]
\centering
\item \textbf{Network position of woman-woman collaborations.}
\end{enumerate}
\vspace{-6pt}
\begin{tabular}{ccc}
\includegraphics[width=0.28\textwidth]{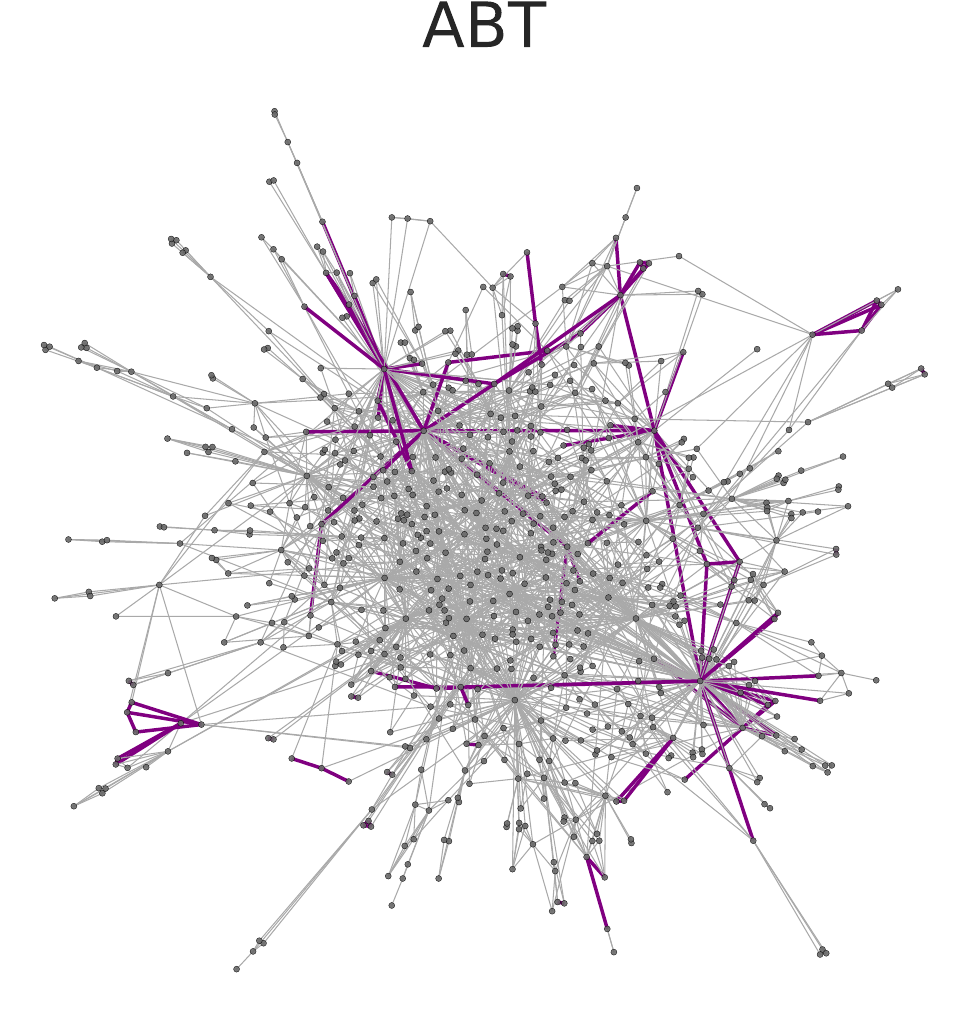} & \includegraphics[width=0.28\textwidth]{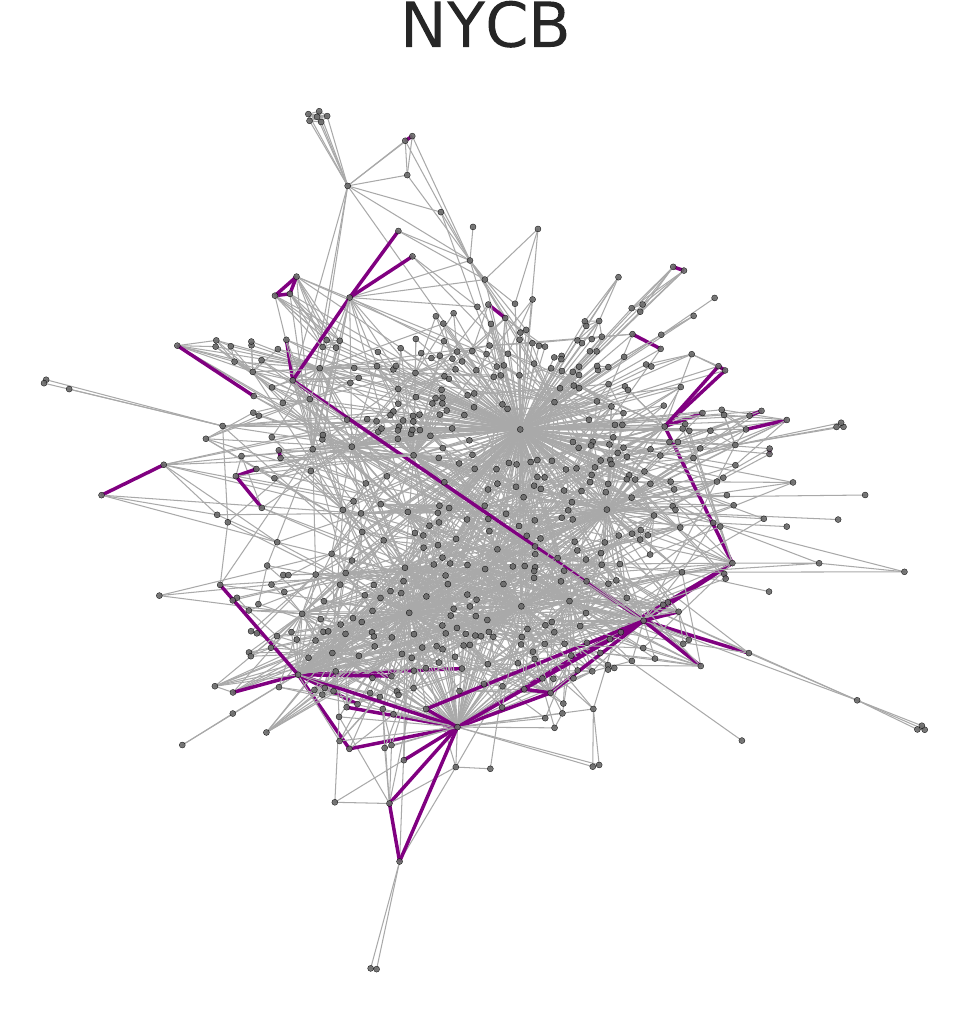} & \includegraphics[width=0.28\textwidth]{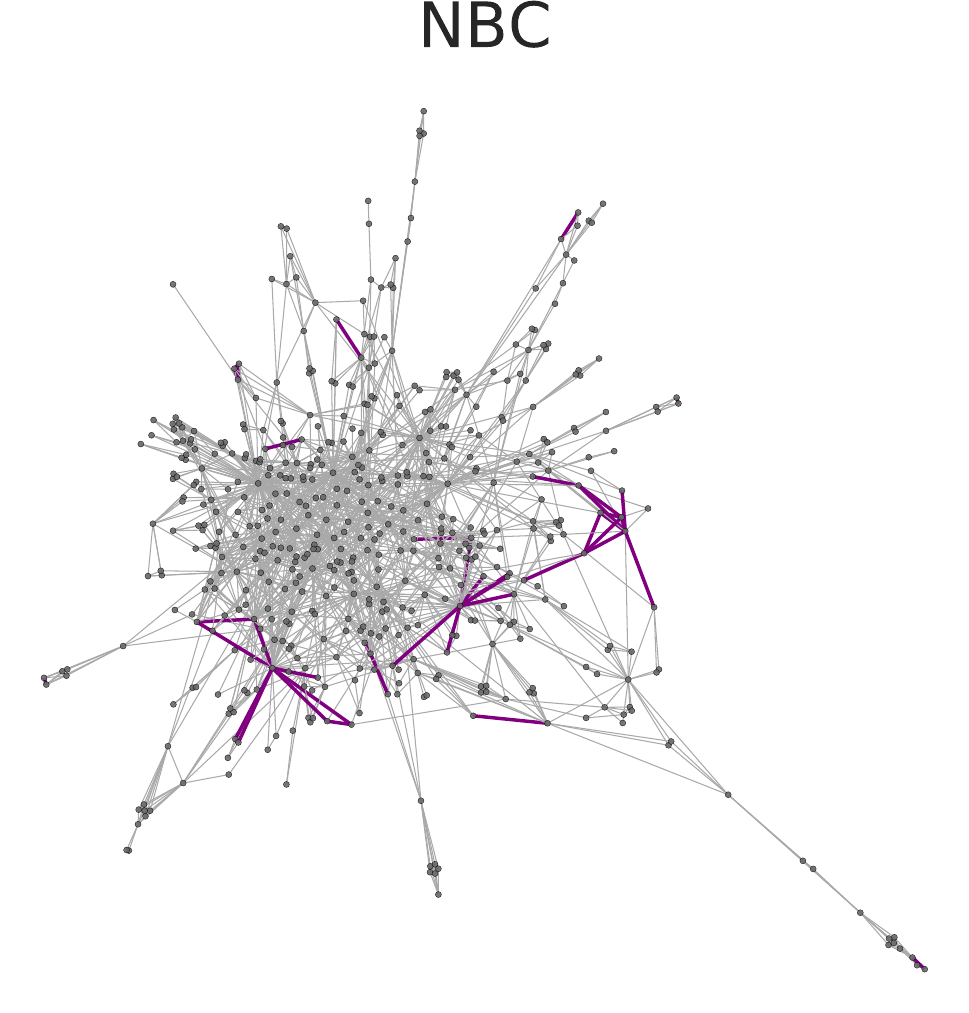} \\
\end{tabular}
\refstepcounter{SIfig}\label{Sfig_supp:networks}
\end{figure}
%%%%%%%%%%%%%%%%%%%%%%%%%%%%%%%%%%%%%%%%%%%%%%%%%%%%%

%%%%%%%%%%%%%%%%%%%%%%%%%%%%%%%%%%%%%%%%%%%%%%%%%%%%%
\begin{figure}[ht!]
\caption{\textbf{Fraction of women of $\text{TCA}$.} Mean of fraction of women, $R_{Women}$ with error bars comparing the empirical networks and the $R_{Women}$ obtained from null models.}
\centering
\includegraphics[width = 0.95\textwidth]{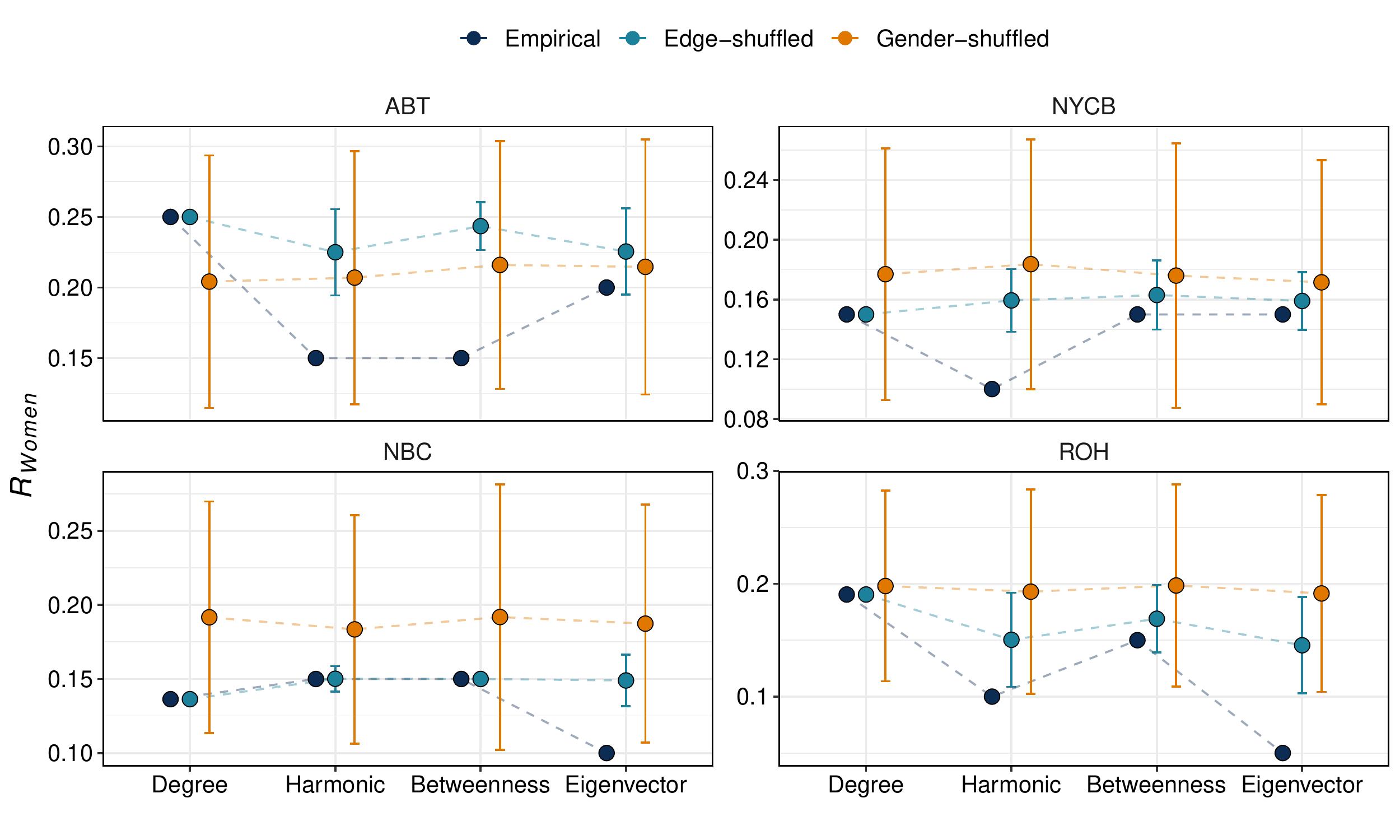}
\refstepcounter{SIfig}\label{Sfig_supp:women_fraction}
\end{figure}
%%%%%%%%%%%%%%%%%%%%%%%%%%%%%%%%%%%%%%%%%%%%%%%%%%%%%

%%%%%%%%%%%%%%%%%%%%%%%%%%%%%%%%%%%%%%%%%%%%%%%%%%%%%
\begin{figure}[ht!]
\centering
\caption{\textbf{$Z$-score of $C(r)_{real}$.} 
Boxplot with error bars. 
Color code is indicated in each panel. 
The red line indicates a $Z$-score equal to zero, corresponding to the mean obtained from each null model.
Panel \textbf{a} shows the distribution of $Z(C)$ for each centrality by gender group $\text{TCA}_{Women, Men}$.
Upper boxes display the Edge-shuffled model, lower boxes the Gender-shuffled model.
Panel \textbf{b} shows the distribution of $Z(\Delta C)$, the Z-score of the gender gap in centrality by rank, separated by null model.}
\includegraphics[scale=0.6]{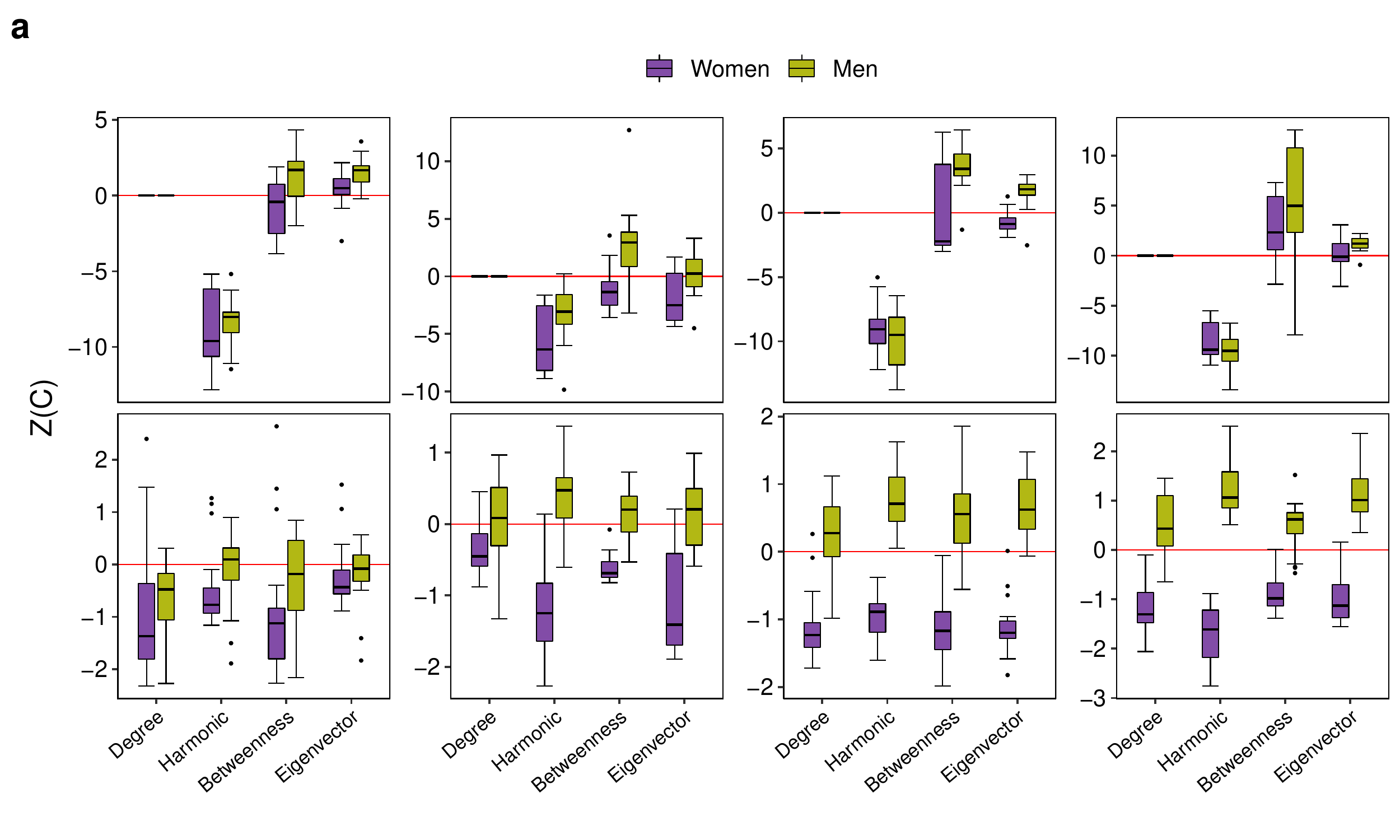}\\
\includegraphics[scale=0.6]{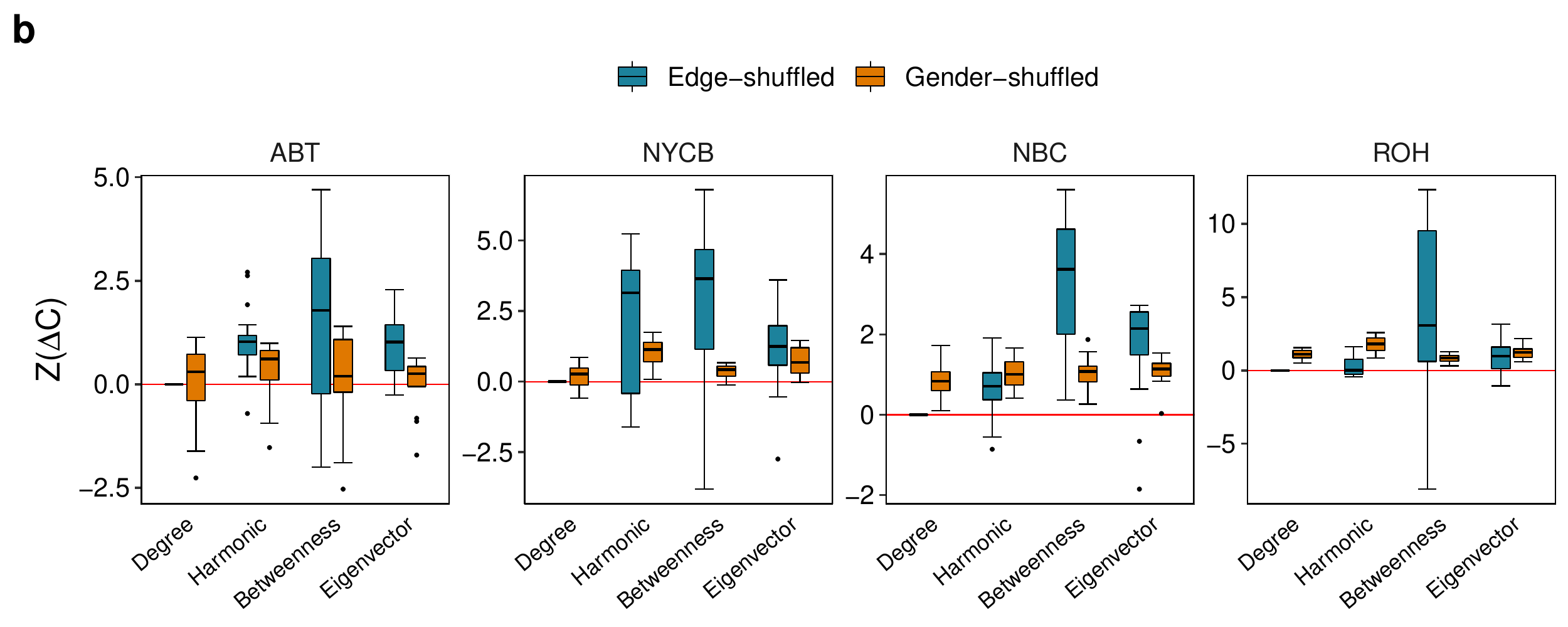}
\refstepcounter{SIfig}\label{Sfig_supp:zscore}
\end{figure}
%%%%%%%%%%%%%%%%%%%%%%%%%%%%%%%%%%%%%%%%%%%%%%%%%%%%%

%%%%%%%%%%%%%%%%%%%%%%%%%%%%%%%%%%%%%%%%%%%%%%%%%%%%%
\begin{figure}[ht!]
\centering
\caption{\textbf{Gender gap in centrality of $\text{TCA}$.} 
Reporting normalized values by company. 
Artists matched by their rank computed in independent gender groups (see Methods).}
\includegraphics[width = 0.95\textwidth]{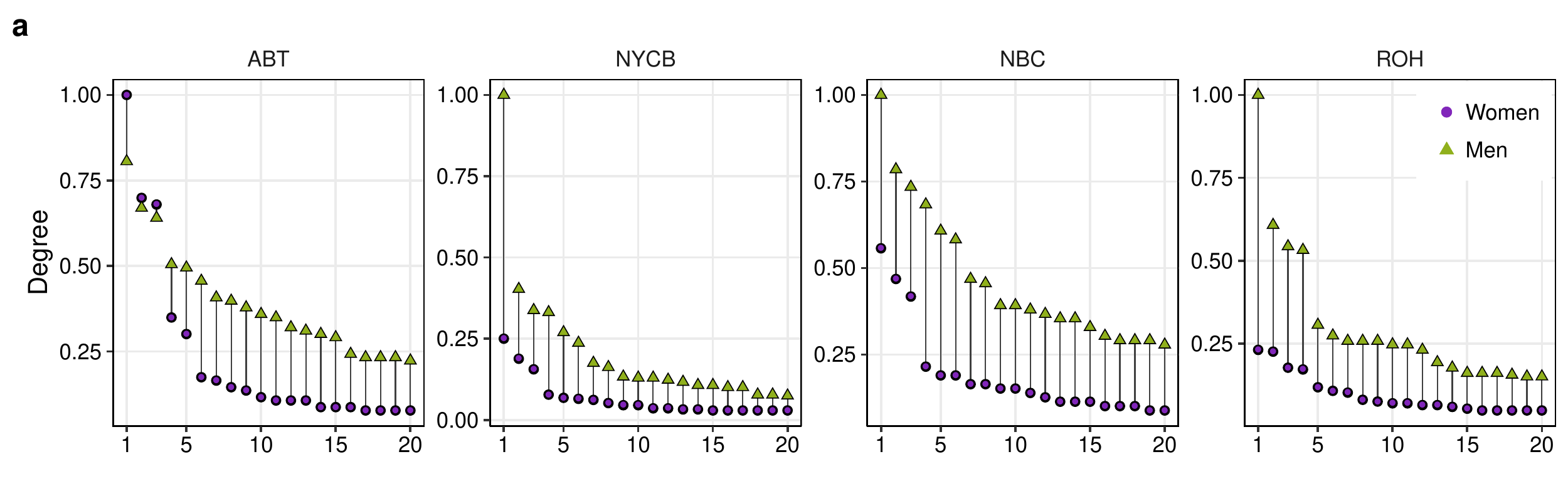}\\
\includegraphics[width = 0.95\textwidth]{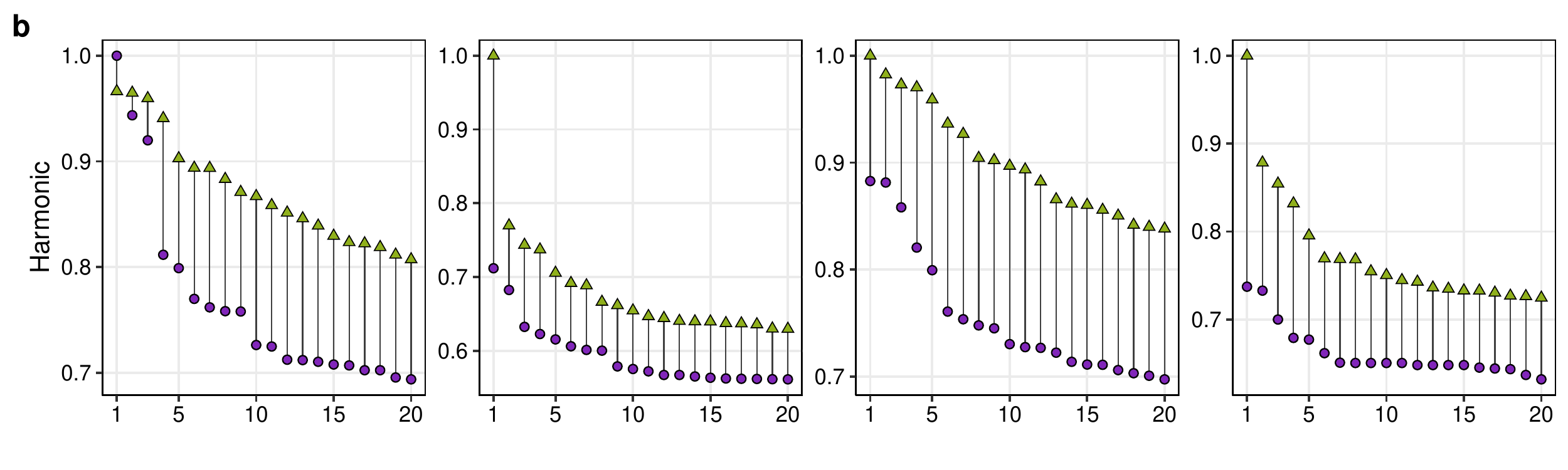}\\
\includegraphics[width = 0.95\textwidth]{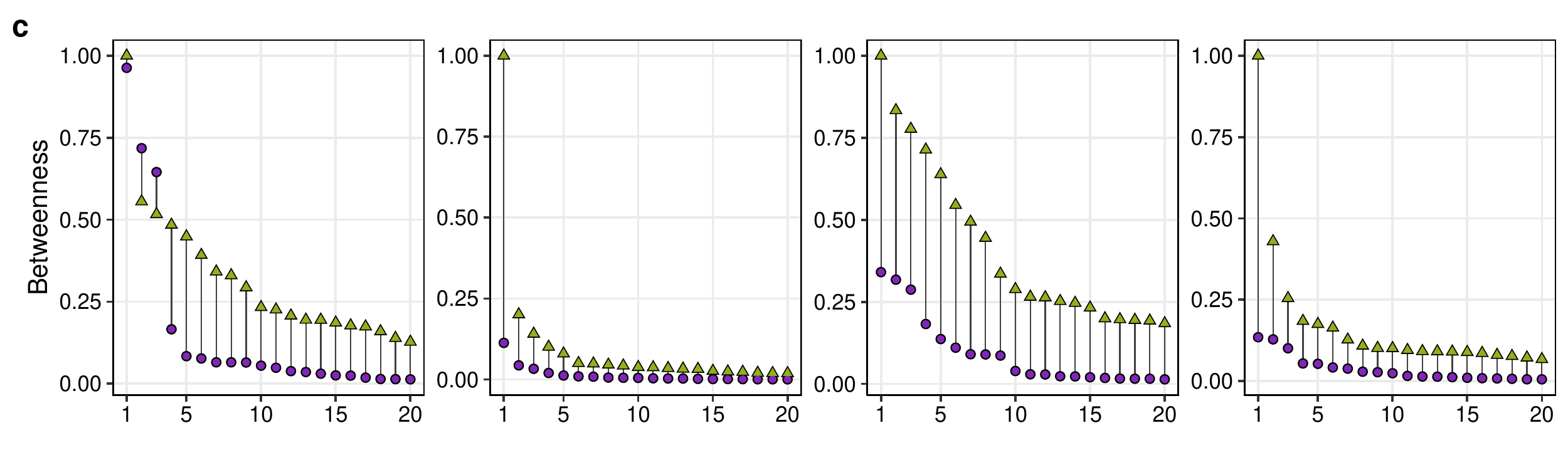}\\
\includegraphics[width = 0.95\textwidth]{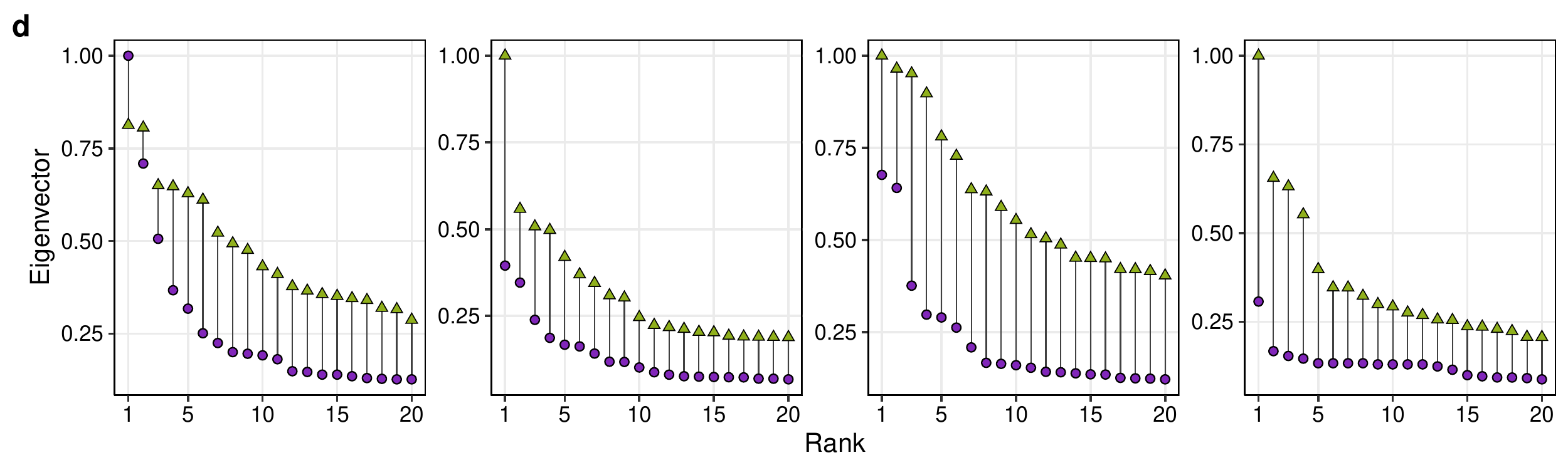}\\
\refstepcounter{SIfig}\label{Sfig_supp:centrality_gap}
\end{figure}
\clearpage
%%%%%%%%%%%%%%%%%%%%%%%%%%%%%%%%%%%%%%%%%%%%%%%%%%%%%

\end{document}